\documentclass{aa}  
\usepackage{multirow}
\usepackage{adjustbox}
\usepackage{graphicx}
\usepackage{relsize}
\usepackage{longtable}
\usepackage{tabularx}
\usepackage{txfonts}
\begin{document}

   \title{Unravelling the complex structure of the Fe~II emission region in Type 1 active galactic nuclei}

   \author{Jelena Kova\v cevi\' c-Doj\v cinovi\'c
   \inst{1},
         Ivan Doj\v cinovi\'c
         \inst{2}        
          \and
         Luka \v C. Popovi\'c
          \inst{1,3}
          }

   \institute{Astronomical Observatory, Volgina 7, 11160 Belgrade, Serbia\\    
     \email{jkovacevic@aob.bg.ac.rs, lpopovic@aob.bg.ac.rs}
   \and
   Faculty of Physics, University of Belgrade, Studentski Trg 12, 11000 Belgrade, Serbia\\
              \email{ivan.dojcinovic@ff.bg.ac.rs}
         \and           
       Department of Astronomy, Faculty of Mathematics, University of Belgrade, Studentski Trg 16, 11000 Belgrade, Serbia\\
       }



  \abstract
  {}
 {Using a large sample of Type 1 active galactic nucleus spectra, we investigated the complex structure of the Fe~II emission region while trying to understand the atomic processes responsible for the enhancement of the Fe~II emission. We explored correlations between Fe~II features and other spectral parameters, with special focus placed on the quasar main sequence, whose underlying physics is crucial for understanding the origin of the strong Fe~II emission.}
{The Fe~II emission was modelled using the flexible Fe~II template that decomposes the optical Fe~II lines into several line groups. According to the atomic properties of transitions, the Fe~II lines were divided into inconsistent and consistent groups (Fe~II$_{incons}$ and Fe~II$_{cons}$), while Fe~II$_{cons}$ lines were additionally decomposed into components originating from different parts of the broad-line region (Fe~II$_{ILR}$ and Fe~II$_{VBLR}$). We traced the behaviour of these line groups and components along the quasar main sequence.}
{Anti-correlation between the equivalent width (EW) of Fe~II  and the full width at half maximum of Fe~II appears to be a more fundamental relation underlying the quasar main sequence. The increase in the EW Fe~II  for smaller line widths is primarily caused by the  strengthening of the EW Fe~II$_{incons}$ lines and, with a smaller contribution, by the enhancement of the EW Fe~II$_{ILR}$  components, while the EW of Fe~II$_{VBLR}$, on average, does not significantly change along the quasar main sequence. The results indicate a possible stratification of the Fe~II emission region occurring in sources with strong Fe~II emission. An increased Eddington ratio may modify the broad-line region structure, leading to the formation of localised regions with specific physical conditions suitable for triggering additional atomic processes. This may result in the appearance of Fe~II$_{incons}$ lines and Fe~II$_{ILR}$ components, which consequently increase the optical Fe~II strength.}
 {}

    \titlerunning{Unravelling the complex structure of the Fe~II emission region in Type 1 AGNs}
               \authorrunning{Kova\v cevi\' c-Doj\v cinovi\'c et al.}                                      

   \maketitle
%

\section{Introduction}\label{Sec1}

The spectra of Type 1 active galactic nuclei (AGNs) exhibit remarkable diversity in the strength and shape of Fe~II emission features, reflecting the complex structure and physical conditions within the broad-line region (BLR).
Numerous correlations have been reported between the Fe~II emission lines and other spectral parameters, although their physical origin remains unclear. Among the most significant and extensively studied is the  anti-correlation between the Fe~II/H$\beta$ ratio and the width of the H$\beta$ line, which was first observed in several early works \citep{Gaskell1985, Zheng1990, Zheng1991}.

\cite{Boroson1992} applied the principal component analysis (PCA) to various observed spectral features, and eigenvector 1 (EV1) of their PCA revealed a significant relationship between several spectral parameters. Among the strongest and the most intriguing correlations of EV1 are the aforementioned anti-correlation between the Fe~II/H$\beta$ ratio (hereafter R$_{FeII}$) and the full width at half maximum (FWHM) of H$\beta$ as well as the anti-correlation between R$_{FeII}$ and the equivalent width (EW) of the  [O~III]$\lambda\lambda$ 4959, 5007 \AA \ lines \citep{Boroson1992}. \cite{Sulentic2000a} proposed that the observed diversity among quasars can be organised in the so-called quasar main sequence, which in the optical plane is represented by the R$_{FeII}$ - FWHM H$\beta$ anti-correlation  \citep{Sulentic2000b, Marziani2001, Sulentic2007,Marziani2018}. Based on the characteristic distribution of sources in the R$_{FeII}$--FWHM H$\beta$ anti-correlation, \cite{Sulentic2000a, Sulentic2000b} introduced the concept of two distinct quasar populations, A and B, separated at the FWHM H$\beta$=4000 km s$^{-1}$. It has been suggested that this division represents a real physical dichotomy among AGNs \citep{Sulentic2011}. 
The underlying physical processes responsible for the R$_{FeII}$--FWHM H$\beta$ anti-correlation have been debated extensively over the past decades. \cite{Boroson1992} proposed that EV1 in their PCA and accordingly the R$_{FeII}$--FWHM H$\beta$ anti-correlation are driven predominantly by the Eddington ratio. Several later studies concluded that the underlying physics of the EV1 is primarily governed by the Eddington ratio in combination with orientation \citep{Sulentic2000a, Marziani2001, Shen2014}.  
 There have been many attempts to theoretically reproduce and model the R$_{FeII}$--FWHM H$\beta$ anti-correlation using spectral synthesis code CLOUDY \citep{Panda2018,Panda2019a,Panda2019b,Floris2024}. Results from modelling indicate that this anti-correlation depends on a combination of many physical parameters, including the Eddington ratio, black hole mass, BLR density and metallicity, micro-turbulence, and orientation \citep[see][]{Panda2024}. 

The most intriguing objects are located at the extreme end of Population A in the quasar main sequence, showing narrow H$\beta$ lines, and the strongest R$_{FeII}$ (the so-called extreme population A) \citep[see][]{Marziani2025}. These very strong Fe~II emitters exhibit several important spectral characteristics that are not yet fully understood and remain difficult to reproduce within the framework of established BLR models. In these objects, the observed ratio of the optical Fe~II to the UV Fe~II lines is significantly higher than theoretically predicted, and several excitation mechanisms and atomic processes have been proposed to explain it \citep[see e.g.][]{Joly1981,Sigut1998,Collin2000}. The model assuming a large optical depth and high column density for the UV Fe~II lines provides the closest agreement with the observed values of this ratio \citep{Collin1980,Joly1981}, suggesting that self-absorption of the UV Fe~II lines has an important role in the enhancement of the optical Fe~II emission relative to the UV Fe~II. One of the proposed explanations for the excess of optical Fe~II emission observed in these spectra is an enhanced iron abundance in the BLR, corresponding to super-solar metallicity \citep{Baldwin2003, Negrete2012, Panda2019a, Sniegowska2021}.

Reproducing the observed optical-to-UV Fe~II ratio as well as the relative intensities among optical Fe~II lines in strong Fe~II emitters remains challenging even for the most advanced photoionisation models that incorporate extensive atomic datasets \citep{Zhang2024,Pandey2025}.
However, in most cases,  modelling studies of strong Fe~II emitters assume that the Fe~II lines originate in a single emission region with specific physical parameters without any stratification. Possible stratification in the Fe~II emission region was investigated by \cite{Popovic2023}, who constructed a sample of model spectra assuming a two-component BLR consisting of two layers with different physical properties: a very broad line region (VBLR) located closer to the black hole and an intermediate-line region (ILR) located further away from the black hole. This sample of synthetic spectra with different ILR and VBLR contributions in H$\beta$ and Fe~II lines succeeded in reproducing the quasar main sequence, where both parameters in this anti-correlation (R$_{FeII}$ and FWHM H$\beta$) were found to depend strongly on the relative contributions from the ILR and VBLR.

The spectra with strong and narrow Fe~II lines have one more particularity apart from the flux excess of optical Fe~II relative to UV Fe~II lines. The relative intensities among optical Fe~II lines do not follow theoretically calculated values \citep{Veron-Cetty2004, Zhang2024}. Namely, some optical Fe~II lines that are expected to be very weak and hardly visible in the spectra according to their transition probabilities in strong Fe~II emitters can be up to two orders of magnitude stronger than predicted. Their observed intensities become comparable to those of other optical Fe~II lines with much higher transition probabilities \citep[see][]{Kovacevic2025}.
 \cite{Kovacevic2025} investigated these optical Fe~II lines and referred to them as inconsistent Fe~II lines (Fe~II$_{incons}$), alluding to the inconsistency between their measured intensities in the spectra and the theoretically expected values. The Fe~II lines whose relative intensities follow the theoretically predicted values were referred to as consistent Fe~II lines (Fe~II$_{cons}$). They found that as the Eddington ratio increases and as the line widths decrease, the intensity of the Fe~II$_{incons}$ lines increases relative to Fe~II$_{cons}$, and at the same time, the total optical Fe~II emission increases relative to UV Fe~II emission.
 They also found that some of the Fe~II$_{incons}$ lines are narrower than Fe~II$_{cons}$ and  suggested that the Fe~II spectrum is likely a complex mixture of radiation originating from emission regions with different physical conditions and distances from the black hole.

All of these results indicate that optical Fe~II lines have a complex origin and should be dissected in order to understand their role in the quasar main sequence and in the other unexplained correlations. Therefore, in this study, we applied the complex Fe~II template presented in \cite{Kovacevic2025}  to analyse the properties of the optical Fe~II lines in a large sample of AGN spectra. This template decomposes the optical Fe~II lines into several line groups according to the atomic properties of their transitions, thus allowing for a separate analysis of the inconsistent and consistent Fe~II lines. We followed the variations in the relative intensities between the Fe~II$_{incons}$ and Fe~II$_{cons}$ lines along the quasar main sequence as well as  changes in the contribution of the Fe~II emission from the ILR and VBLR. In this way, we sought to obtain a better understanding of the physical processes responsible for the strong Fe~II emission and to explain the physics underlying the quasar main sequence. 

In Sect. 2, we describe our sample and the method of spectral decomposition. The results are presented in Sect. 3 and discussed in Sect. 4. Finally, our conclusions are summarised in Sect. 5.

\section{Sample and analysis}\label{Sec2}

\cite{Kovacevic2025} used the sample of 1046 Type 1 AGN spectra for investigation of the Fe~II emission lines and atomic processes involved in their emission. Since that sample is well defined and chosen to have high-quality spectra convenient for sophisticated spectral decomposition and analysis, we used the same sample for this research. Detailed description of the sample selection, spectral correction for reddening, redshift and host galaxy contribution, as well as fitting procedure, is given in \cite{Kovacevic2025}. Here we list the most important steps in that process.

The sample is taken from Sloan Digital Sky Survey (SDSS) database, Data Release 16 (DR16) \citep{Ahumada2020} using structural query language (SQL) in order to satisfy the following criteria: objects to be Type 1 AGNs, cosmological redshift (z) to be z < 0.7 and with no redshift warning, and median S/N per pixel of the whole spectrum and g-band to be greater than 30. In this way we got the high-quality spectra which cover optical Fe~II emission in the 4000-5600 \AA \ range, in which Fe~II lines can be precisely fitted. For correction of the Galactic extinction, we used the standard extinction law given in \cite{Howarth1983} and extinction coefficients given in \cite{Schlegel1998}. After correction for the cosmological redshift, we decomposed the spectra to the host and pure-AGN contribution using the method of the spectral PCA, as it is described in \cite{VandenBerk2006}. In this method we fitted the spectra with the linear combination of two independent sets of eigenspectra, derived from the  pure-quasar \citep{yip04a} and the pure-galaxy \citep{yip04b} SDSS samples. The host galaxy contribution is obtained as the linear combination of only pure-galaxy eigenspectra, obtained from the best fit. The pure-AGN spectra are obtained after subtraction of the host galaxy contribution from the observed spectra. The example of the spectral decomposition to the host galaxy and the pure-AGN contribution could be seen in \cite{Kovacevic2025} (see their Fig. 1).

\begin{figure}
	\centering
	\includegraphics[width=90mm]{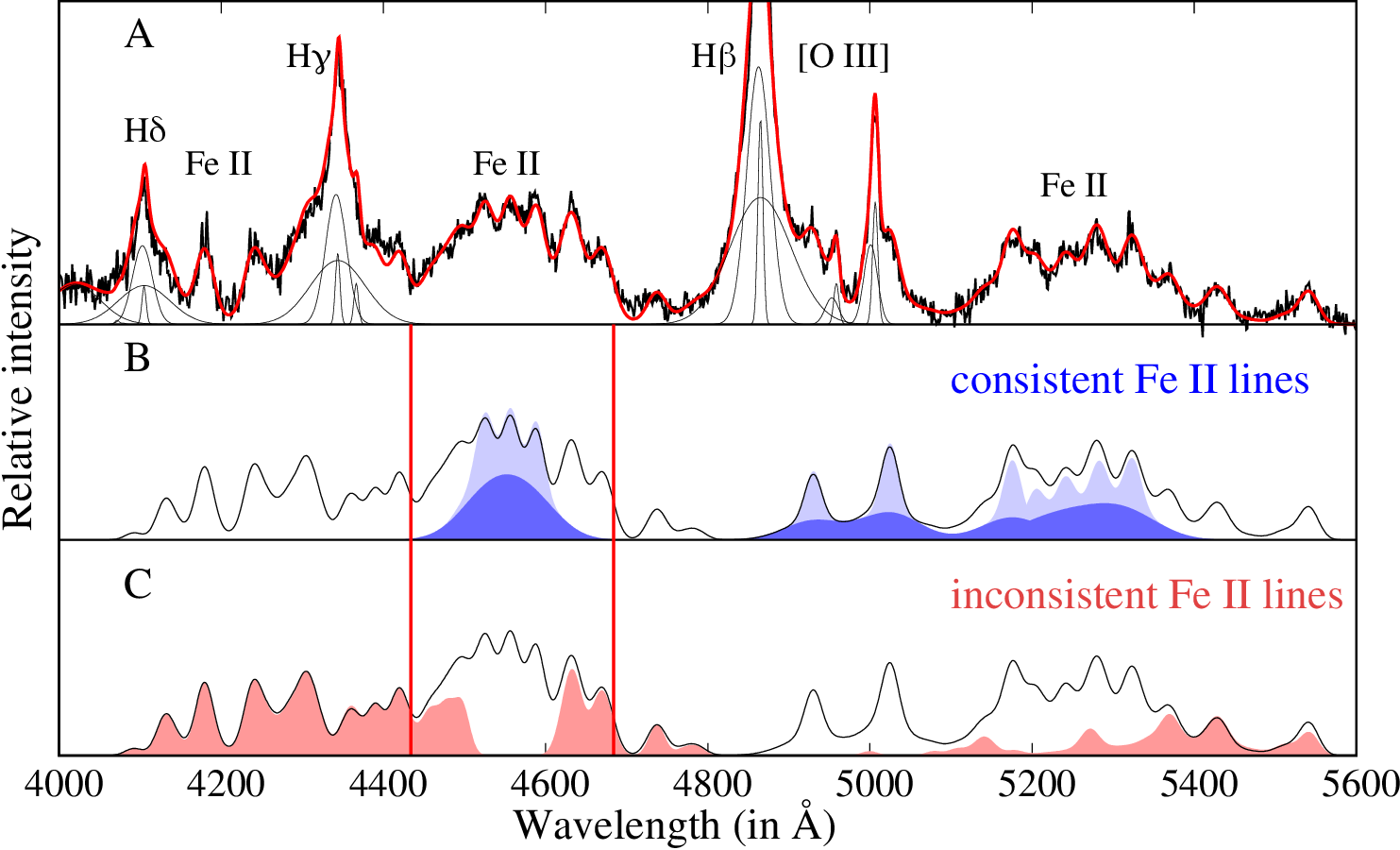}
	      \caption{Example of the spectral decomposition in the 4000-5600 \AA \ range. Panel A shows the multi-Gaussian decomposition for SDSS J111941.12+595108.7, where the best fit is denoted with a red line. The Fe~II template is shown separately as black solid line in Panels B and C. In Panel B, the consistent Fe~II lines are shaded with the blue colour. The sum of the ILR components is coloured with light blue, and sum of the VBLR components in dark blue. In Panel C, the inconsistent Fe~II lines are coloured with red. Two vertical red lines in Panels B and C denote the spectral range of 4434-4684 \AA, usually used for calculation of the R$_{FeII}$. }
    \label{fig00}
\end{figure}

Afterwards, the pure AGN spectra were fitted with the multi-Gaussian model. In applied fitting procedure all emission lines in 4000-5600 \AA \ range and continuum emission were fitted simultaneously. The continuum is modelled with power-law, while emission lines were modelled with single or multiple Gaussian functions. The Balmer lines in 4000-5600 \AA \ range were fitted with three Gaussian components: one narrow Gaussian that fits the emission from the narrow-line region (NLR), while emission from BLR is described with sum of the two broad Gaussians, as we assume that broad component of the Balmer lines originate from two layers of the BLR. One is the ILR, which is supposed to be outermost area of the BLR, farther away from the black hole. The emission from this region is represented with Gaussian that fits the core of the broad lines. The other sub-region is VBLR, which is the inner layer of the BLR and is represented with broader Gaussian which fits wings of the lines. This two-Gaussian decomposition approach for BLR emission is commonly used for spectra with very broad Balmer lines \citep[see e.g.][]{Hu2008, Li2015}. However, several studies have shown that Lorentzian profiles provide a good fit of the broad emission-line shapes in the case of the NLSy1 galaxies, and more generally in Population A sources \citep[see e.g.][]{Veron-Cetty2001, Sulentic2002}. We find that the broad Balmer line profiles in NLSy1 galaxies can be also very well reproduced by the sum of two Gaussian components resulting in the 'Lorentzian-like' shape.  We therefore interpret the broad Balmer line shapes in NLSy1s  as the superposition of Gaussian components with different widths, reflecting Doppler broadening in distinct BLR layers, rather than a single Lorentzian profile, which is expected to arise by some other broadening mechanism (collisional broadening, Stark broadening, etc.). This approach allowed us to apply a uniform Balmer-line decomposition model across the entire AGN sample, independent of Balmer line widths, while preserving a consistent physical interpretation of Doppler broadening as the dominant mechanism affecting broad-line AGN profiles.
The widths and shifts of the ILR Gaussians are fixed to be the same for all Balmer lines  within one spectrum. The same is done for the widths and shifts of the VBLR Gaussians and ILR/VBLR intensity ratio. Additionally, the relative intensities of the broad components of H$\beta$, H$\gamma$ and H$\delta$ lines are fixed to follow the theoretical value for case B \citep[see][]{Osterbrock2006}.

The numerous Fe~II lines are fitted with the Fe~II model given in \cite{Kovacevic2025},  which is the improved version  of the Fe~II template given in \cite{Kovacevic2010} (see more details in Sec \ref{Sec2.1}). The narrow lines in 4000-5600 \AA \ range (narrow Balmer components, [O~III] 4959, 5007 \AA, [O~III] 4363 \AA \ and [S II] 4068, 4076 \AA) are fitted with single Gaussian components, with the same widths and shifts. In the case of observed asymmetry in [O~III] 4959, 5007 \AA \ profile, additional wing component is included in the modelling of these lines. The He II 4686 \AA \ and He I 4026 \AA \ lines are fitted with single broad Gaussians. The example of the spectral decomposition in 4000-5600 \AA \ range is given in Figure \ref{fig00}, Panel A.

\subsection{Flexible Fe~II Template:  A tool for investigation of the complex Fe~II emission}\label{Sec2.1}

Several studies have shown that the relative intensities among Fe~II multiplets vary in different AGN spectra \citep{Kovacevic2010, Shapovalova2012, Ilic2023, Kovacevic2025}. Therefore, empirical Fe~II templates with fixed relative intensities among Fe~II lines cannot accurately fit Fe~II emission in the case of spectra with significantly different Fe~II properties \citep[see appendix in][]{Kovacevic2010, Kovacevic2025}.
 The semi-empirical Fe~II model given in \cite{Kovacevic2010} separate the Fe~II lines into several line groups according to atomic properties of transitions, giving them the free parameter of intensity, achieving in this way the flexibility of the model and possibility for analysis of atomic processes. As a result, the template gives the intensities of the different Fe~II line groups, following the 'breathing' of the Fe~II lines in different spectra.  \cite{Kovacevic2025} upgraded and improved the Fe~II model given in \cite{Kovacevic2010} by addressing some deficiencies of initial model and giving more degrees of freedom for fitting the complex Fe~II emission. These modifications result in a more accurate fit of Fe~II lines that can successfully reproduce Fe~II emission of various properties. The great flexibility of this template also allows for investigation of the behaviour of a certain group of Fe~II lines, giving a better insight into the complex physics of Fe~II emission. 
 
 In the Fe~II model given in \cite{Kovacevic2025}, the Fe~II lines in 4000-5600 \AA \ range are divided into two large groups. The first group comprises the lines whose observed relative intensities follow the values obtained by theoretical calculations using their atomic parameters (so-called consistent Fe~II lines, Fe~II$_{cons}$). The other group contains those lines whose observed relative intensities could be significantly larger than theoretically expected. These lines are addressed as inconsistent Fe~II lines (Fe~II$_{incons}$), which alludes to the inconsistency of their measured intensities with theoretically predicted. 
 While the consistent Fe~II lines arise from allowed transitions, the inconsistent Fe~II lines are mostly intersystem and intercombination multiplets, with the addition of several forbidden multiplets, allowed lines and several transitions originating from highly excited levels.  Most of Fe~II$_{incons}$ lines have up to two orders of magnitude lower Einstein coefficients for spontaneous emission comparing the Fe~II$_{cons}$. Consequently, they should be barely detectable in optically thin spectra, even when Fe~II$_{cons}$ lines are very strong. However, in some spectra their intensities are comparable to those of the consistent Fe~II lines, suggesting that additional atomic processes enhance their emission. Most of Fe~II$_{incons}$ lines share the same upper transition level as Fe~II$_{cons}$ lines, while the lower levels of both groups have similar energies and are all metastable. One of the possible mechanisms which may increase intensities of the Fe~II$_{incons}$ lines could be the self-absorption of Fe~II$_{cons}$ lines. Since Fe~II$_{incons}$ lines have significantly lower emission and absorption probabilities  than Fe~II$_{cons}$, and both have lower metastable levels, the process of self-absorption would be more efficient for  Fe~II$_{cons}$ lines and would lead to a redistribution of energy from  Fe~II$_{cons}$  to the Fe~II$_{incons}$ lines, thereby increasing the intensities of Fe~II$_{incons}$ \citep[see][]{Kovacevic2025}.

 The consistent Fe~II lines were divided into three line groups (F, S, and G) formed following their lower term of transition. The intensities of the line groups were set to be the free parameters, while the relative intensities among the Fe~II lines within these groups were calculated using formula 1 in \cite{Kovacevic2010}. The widths of the Fe~II lines are strongly correlated with those of the broad Balmer lines \citep{Kovacevic2010, Kovacevic2015}, indicating important role of  Doppler broadening in shaping of their profiles. Therefore, we initially applied the same two-component approach used for the broad Balmer lines to model the Fe~II line profiles across the entire sample. The consistent Fe~II lines, which are the strongest in Fe~II bump, were fitted with a two-Gaussian model, where the ILR component fits the core and the VBLR component fits the wings of the lines, assuming that these components arise in kinematically distinct layers of the BLR.  We assumed that the widths of ILR and VBLR components as well as their intensity ratio are the same for all Fe~II consistent lines within one spectrum. None of these parameters were tied with the same parameters of the Balmer lines in the fitting procedure. In Appendix \ref{A}, we analyse the justification for using a two-component model for fitting of Fe~II$_{cons}$ lines, and we found that this approach provides a significantly better fit for spectra with narrower Fe~II lines, where individual line peaks are visible within the Fe~II bump. However, for spectra with very broad Fe~II lines, with approximately FWHM Fe~II > 5000 km s$^{-1}$, its application does not improve the fit (see Figure \ref{fig001}). In these spectra, the Fe~II lines form smooth bumps without distinguishable Fe~II line peaks, so both the ILR and VBLR Gaussians overlap, as they have very broad and similar widths. This is not in accordance with  the assumptions of the two-component model, which predicts that the ILR components have significantly narrower widths than the VBLR components, as they originate from kinematically different regions. Therefore, the two-component Fe~II model was applied only for spectra with estimated width of Fe~II lines smaller than 5000 km s$^{-1}$. For spectra with FWHM Fe~II > 5000 km s$^{-1}$, the Fe~II emission lines were fitted with a single Gaussian model for each Fe~II line, assuming that all Fe~II emission arises from a kinematically uniform emission region with very high Doppler velocities. An example of the Fe~II decomposition for a spectrum with FWHM Fe~II < 5000 km s$^{-1}$, where consistent Fe~II lines are fitted with two-component model, is shown in Figure \ref{fig00}, Panel B. 

The inconsistent Fe~II lines were divided into four line groups (P+, G+, H and OL), following the atomic characteristic of their transitions and similar behaviour in spectra. Each of the line groups has the free parameter of intensity, while the relative intensities among the lines within groups were fixed since they are determined empirically \citep[see][]{Kovacevic2025}. The inconsistent Fe~II lines were fitted with single Gaussian model, and each group has a free parameter for width. The shift was taken to be the same for all Fe~II lines in model. The inconsistent Fe~II lines are highlighted in Figure \ref{fig00}, Panel C.

To summarise, in the complex Fe~II model, the intensities of consistent lines are described with five parameters. These are three parameters of intensity for F, S, and G line groups, temperature for calculation of the relative intensities of lines within the groups, and one parameter for intensity ratio of VBLR/ILR components. The other free parameters that describe the consistent Fe~II lines are the widths of ILR and VBLR components and the shift. The inconsistent Fe~II lines are described with free parameter of intensity and width for each of line groups (P+, G+, OL and H). All consistent and inconsistent Fe~II line groups with a free intensity parameter have at least some lines that do not overlap with the lines from the other line groups. This enables a reliable determination of their fluxes and reduces the possibility of flux redistribution among them due to fitting degeneracies. Detailed explanation of complex Fe~II model, with summarised properties and list of the included lines in different line groups, is given in  \cite{Kovacevic2025} (see their Appendix B).

\subsection{Measurements of spectral properties and error estimates}\label{Sec2.2}

For purpose of this research we measured FWHMs, EWs and fluxes of H$\beta$ and Fe~II lines. The FWHM of the broad H$\beta$ is measured using the sum of the ILR and VBLR H$\beta$ components, obtained from the best fit. Maximum of the sum is normalised to the unit, and the width at half intensity is measured. The dispersion of the broad H$\beta$ line was computed as the  flux-weighted second moment of the total broad line profile \citep{Peterson2004}.

\begin{figure}
	\centering
\includegraphics[width=65mm]{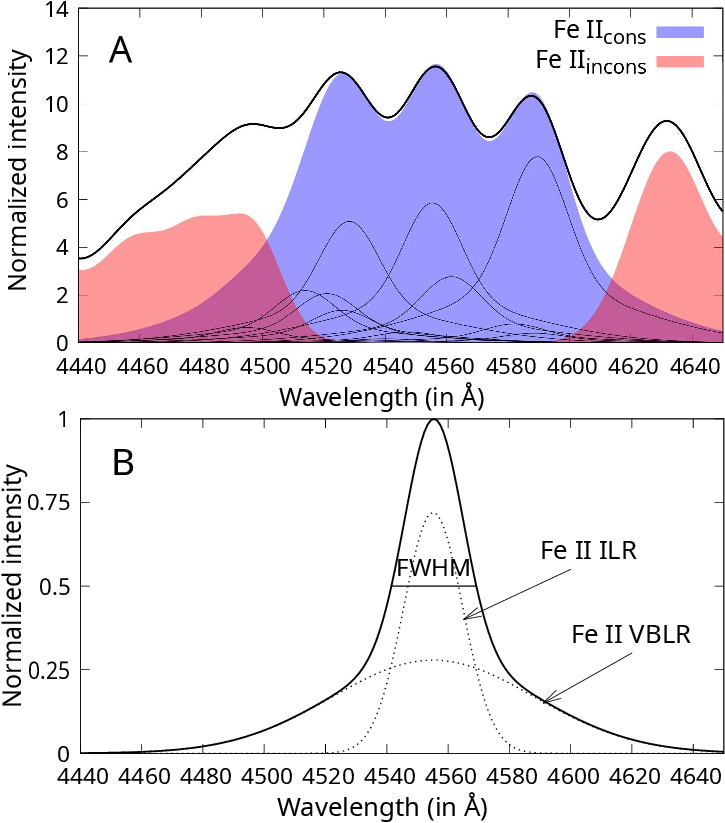}
 \caption{Measuring the FWHM of Fe~II$_{cons}$ lines fitted with the two-Gaussian model. 
Panel A:  Sum of Fe~II$_{cons}$ lines. The Fe~II$_{cons}$ lines are shown with a solid thin lines, while their sum is coloured in blue. Panel B: Single Fe~II$_{cons}$ line. We extracted one Fe~II$_{cons}$ line ($\lambda$4549.5 \AA), which consists of the sum of the ILR and VBLR components, and measured the FWHM of the entire line profile. }
  \label{fig010}
\end{figure}

Since the Fe~II$_{cons}$ are also fitted with two-Gaussian model for spectra where  Fe~II width is smaller than 5000 km s$^{-1}$, their FWHM is measured in the same manner as for H$\beta$. We singled out one Fe~II consistent line, which is the sum of the ILR and VBLR Fe~II components. We normalised it to  unity and measured the width at half maximum of its overall shape as  illustrated in Figure \ref{fig010}. The FWHMs of the inconsistent Fe~II line groups are obtained directly from the best fit, as well as the widths of the consistent Fe~II  lines broader than 5000 km s$^{-1}$, which are fitted with a single-Gaussian model. 
The consistent Fe~II lines are present in all objects with observed Fe~II lines, which is 1022 objects from total sample, while  majority of inconsistent Fe~II lines disappear in spectra with broader emission lines (FWHM > 5000 km s$^{-1}$) \citep[see][]{Kovacevic2025}. Therefore, in this work we use only the measured width of the consistent Fe~II lines, in further text assigned as FWHM Fe~II. The flux and EW of H$\beta$ are measured for a total broad line. In the case of Fe~II, the flux and EW are measured for the sum of the multiple lines: all consistent Fe~II (Fe~II$_{cons}$), all inconsistent Fe~II (Fe~II$_{incons}$) and for all Fe~II lines in 4000-5600 \AA \ range (Fe~II$_{cons}$+Fe~II$_{incons}$), denoted as Fe~II$_{tot}$. In the case of the  Fe~II$_{cons}$, the fluxes and EWs are also separately measured for sum of their ILR (Fe~II$_{ILR}$) 
and VBLR (Fe~II$_{VBLR}$) components (see Figure \ref{fig00}, Panel B).

For accurate measuring of the fluxes and specially EWs of the lines, the correct determination of the continuum level is very important. \cite{Popovic2023} found that in the case of spectra with strong and very broad Fe~II lines, the optical Fe~II quasi-continuum can be very strong, and it is difficult to separate it from the power-law continuum. It can cause the effect of underestimated fluxes of the Fe~II and H$\beta$ lines and slightly overestimated flux of continuum under these lines. Consequently, the most affected measured parameters are the EWs, especially EW of Fe~II, which may be strongly underestimated, in the case of the strong Fe~II pseudocontinuum.
In order to reduce the uncertainty in the determination of continuum level  caused by Fe~II pseudocontinuum, we have taken several steps in fitting procedure \citep[see Appendix A in][]{Kovacevic2025}. The most important steps are simultaneously fitting of the continuum power-law  with the emission lines, fixing the relative intensities of the Balmer lines to follow theoretical values for case B, and including of two-Gaussian model for consistent Fe~II lines in the case of sources where the application of that model is justified.  The EWs of H$\beta$ and Fe~II lines are measured relative to continuum under these lines. 

The bolometric luminosity (L$_{bol}$) was calculated as $L_{bol} = 9\cdot\lambda L_{5100}$ \citep{Kaspi2000} and the Eddington luminosity (L$_{Edd}$) as $L_{Edd}=1.26\cdot10^{38}(M_{BH}/M_{sun})$ erg s$^{-1}$ \citep{Wu2004}. The Eddington ratio (R$_{Edd}$) was estimated as R$_{Edd}$ = L$_{bol}$/L$_{Edd}$. The adopted cosmological parameters for luminosity calculation were $\Omega_M$=0.3, $\Omega_\Lambda$=0.7, and $\Omega_k$=0, and a Hubble constant of $\rm H_{0}$=70 km s$^{-1}$ Mpc$^{-1}$ was assumed.

We estimated the fitting uncertainties using a Monte Carlo method. Since we applied a different Fe~II modelling approaches for spectra with  narrow and very broad  Fe~II lines, we divided the total sample into two subsamples, with the division set at an FWHM Fe~II = 5000 km s$^{-1}$. We calculated the mean spectrum for each subsample by normalising the spectra at the same continuum level at $\lambda$5630 \AA. Then, we created  two subsets of  100 mock spectra by adding random noise to the mean spectra. The level of noise in mock spectra is taken to be as in the spectrum with the highest measured noise in the sample. In this way we wanted to estimate the upper limit of uncertainties. After fitting procedure, we took the 1$\sigma$ dispersion of the parameters as the parameter uncertainty, similarly as done in \cite{Kovacevic2025}. For spectra with an FWHM Fe~II less than 5000 km s$^{-1}$, we estimated the following uncertainties: $\sim$ 5\% for  FWHMs of  H$\beta$ and Fe~II, and up to $\sim$6\%
for all measured fluxes (Fe~II$_{tot}$, Fe~II$_{cons}$, Fe~II$_{incons}$, Fe~II ILR and VBLR components, narrow and broad components of H$\beta$ and [O~III]$\lambda$5007 \AA). The upper limit on the uncertainties in the shifts of the narrow lines and [O~III] wing components is $\sim$ 5\%. For spectra with FWHM Fe~II larger than 5000 km s$^{-1}$, parameter uncertainties are: $\sim$ 7\% for FWHMs of  H$\beta$ and Fe~II, and up to  $\sim$ 7\% for all measured fluxes of lines. The upper error in shift of the narrow lines and [O~III] wing components is $\sim$ 6\%. The uncertainties of the EWs are estimated as 1$\sigma$ dispersion of the measured EWs of the lines for 100 mock spectra, in each subsample. For the subsample with an FWHM Fe~II less than 5000 km s$^{-1}$, we obtained uncertainties up to  $\sim$ 6\% for all measured EWs of lines and their components, and for spectra with an FWHM Fe~II greater than 5000 km s$^{-1}$, we obtained uncertainties up to  $\sim$ 7\%. 

Since we used a two-component model for decomposition of the consistent Fe~II lines for a subsample with an FWHM Fe~II smaller than the threshold of  5000 km s$^{-1}$, we investigated if objects near $\sim$ 5000 km s$^{-1}$  may switch between decomposition models due to noise. For spectra with FWHM Fe~II near the 5000 km s$^{-1}$, the upper limit for  uncertainty is 7\%, corresponding to 5000 $\pm$ 350 km s$^{-1}$. In our sample, 22 objects fall within this range, where placement into one of the two subsamples is uncertain. Since this represents only a small fraction of the total sample of 1022 spectra, we conclude that these uncertainties do not significantly affect our analysis. 

We also investigated possible parameter degeneracies due to noise, testing whether part of the Fe~II flux could be traded against the continuum emission or redistributed among different Fe~II components during the fitting procedure. We searched for correlations between the flux residuals, defined as the difference between the fluxes measured in the mock spectra and those of the input model. We found a moderate anti-correlation between the residuals of the total Fe~II flux and the continuum level for the mock spectra generated from both models (with narrower and with broader Fe~II lines). For the mock spectra generated from the model with narrower Fe~II lines, we also found a moderate anti-correlation between the residuals of the ILR and VBLR Fe~II components. These anti-correlations indicate a partial flux redistribution between these components induced by the noise.  These effects are accounted for in the estimated flux uncertainties, which do not exceed 7\% for the fluxes of different Fe~II components. No correlation is found between the flux residuals of the consistent and inconsistent Fe~II lines.

\section{Results}\label{Sec3}

\subsection{Quasar main sequence: Searching for the underlying correlation }\label{Sec3.1}

We investigated in more details the quasar main sequence  (Fe~II/H$\beta$ versus FWHM H$\beta$ correlation), using the Fe~II parameters obtained from the best fit with complex Fe~II model. The flux ratio of Fe~II and  H$\beta$ in this relationship  is usually assigned as R$_{FeII}$ parameter, and it represents the ratio of the integrated flux of Fe~II in 4434-4684 \AA \ range, and the flux of the broad component of  H$\beta$. The FWHM H$\beta$ is also measured only for the broad component of the line, after removal of the narrow component. Since this relationship is complex, with three spectral parameters included (Fe~II and H$\beta$ fluxes and the H$\beta$ width), we tried to simplify it, and to find the underlying correlation, which can give more clues about the physics of the emission region. Through this analysis, we used the Spearman rank correlation coefficient, which is sensitive to monotonic relationships without assuming linearity and is less affected by outliers, making it appropriate for analysis of non-linear parameter dependencies.

\begin{figure}
	\centering
	\includegraphics[width=60mm, angle=270]{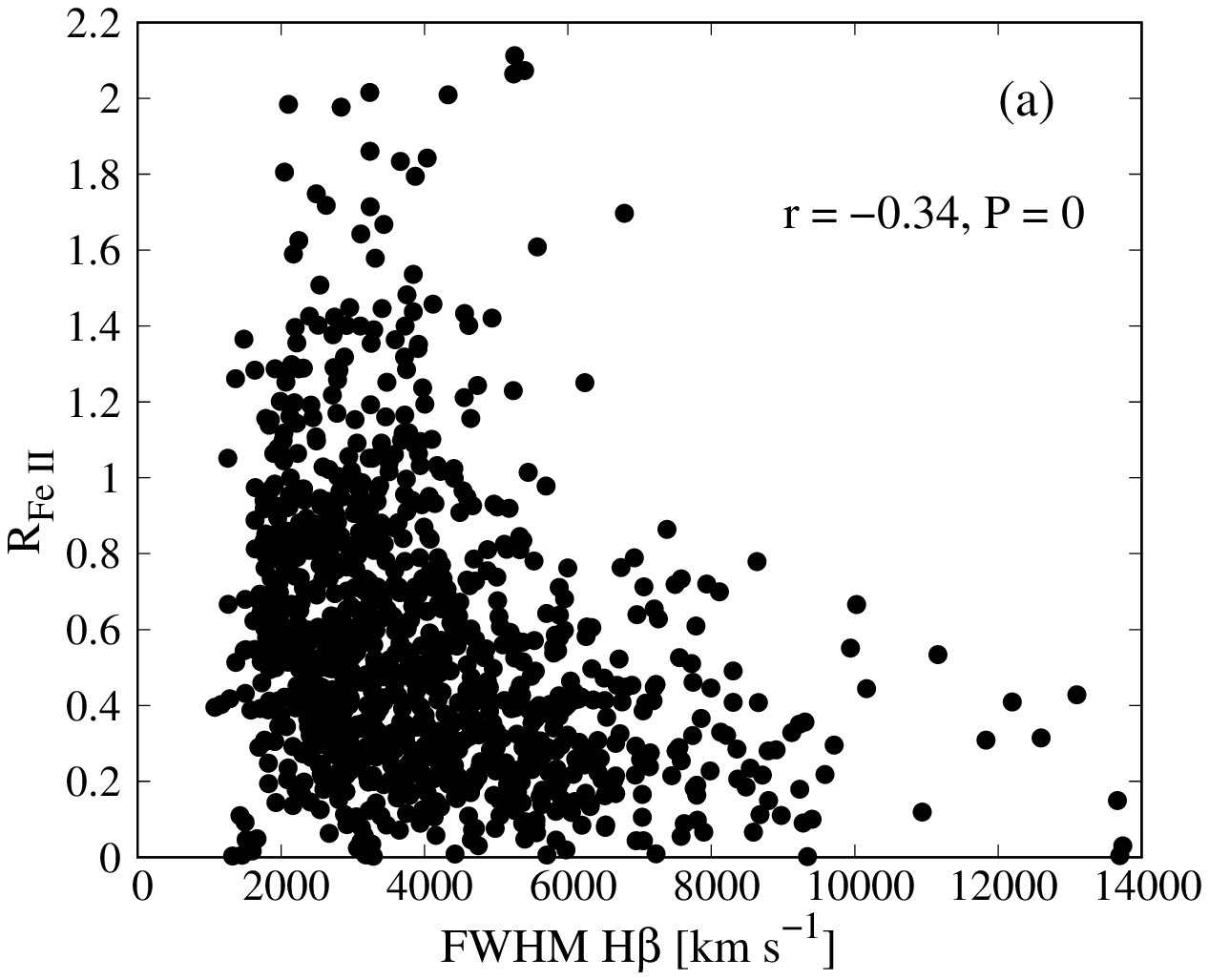}
	\includegraphics[width=60mm, angle=270]{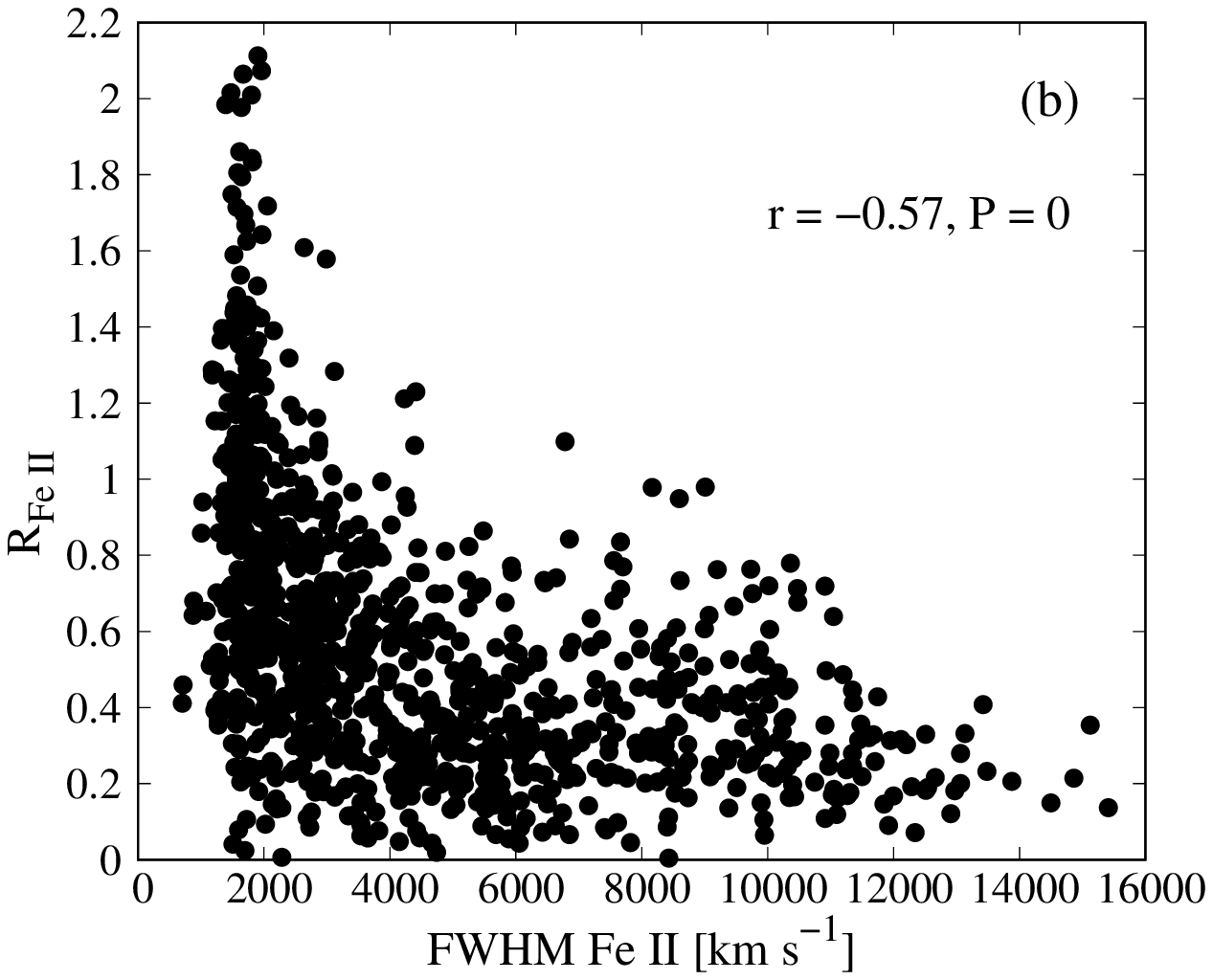}	
       \caption{Relationship between FeII [4434-4684]/H$\beta$ ratio (R$_{FeII}$) and FWHM H$\beta$ (a) and  FWHM Fe~II (b). The Spearman coefficients of correlations (r) and P-values are given in graphs.}
    \label{fig01}
\end{figure}

In Figure \ref{fig01} (a), we plot the quasar main sequence  for our sample. 
As expected, we found the negative correlation between these parameters (r = -0.34, P-value = 0, where r is the Spearman coefficient of correlation).  If we take the larger spectral range of the integrated Fe~II flux (4000-5600 \AA, in further text Fe~II$_{tot}$), instead narrow range of 4434-4684 \AA, the coefficient of correlation does not change. Since our Fe~II template allows the FWHM of the Fe~II lines to be obtained (see Section \ref{Sec2.2}), we tried to plot FWHM Fe~II versus R$_{FeII}$ ratio, instead standardly used FWHM H$\beta$.  We found that anti-correlation between these parameters  is sharper compared to the  FWHM H$\beta$ versus R$_{FeII}$ anti-correlation. The anti-correlation significantly increases and the Spearman coefficient of correlation becomes r = -0.57, P = 0 (see Figure \ref{fig01} (b), Table \ref{T01}). Although the widths of H$\beta$ and Fe~II are highly correlated with each other (r = 0.66, P = 0), it seems that FWHM Fe~II gives the better trace for the underlying physics of this anti-correlation.

\begin{figure}
	\centering
	\includegraphics[width=60mm, angle=270]{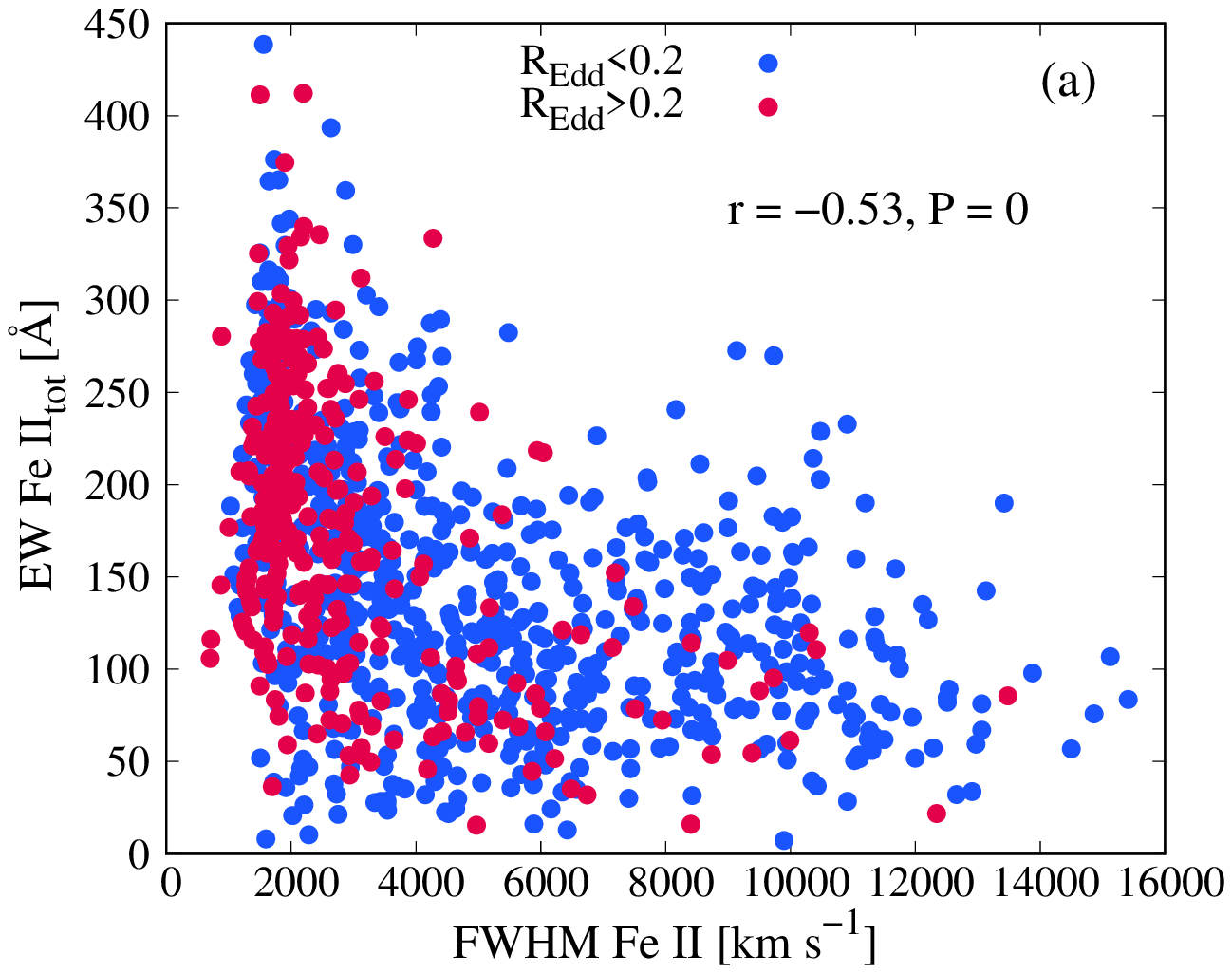}
	\includegraphics[width=60mm, angle=270]{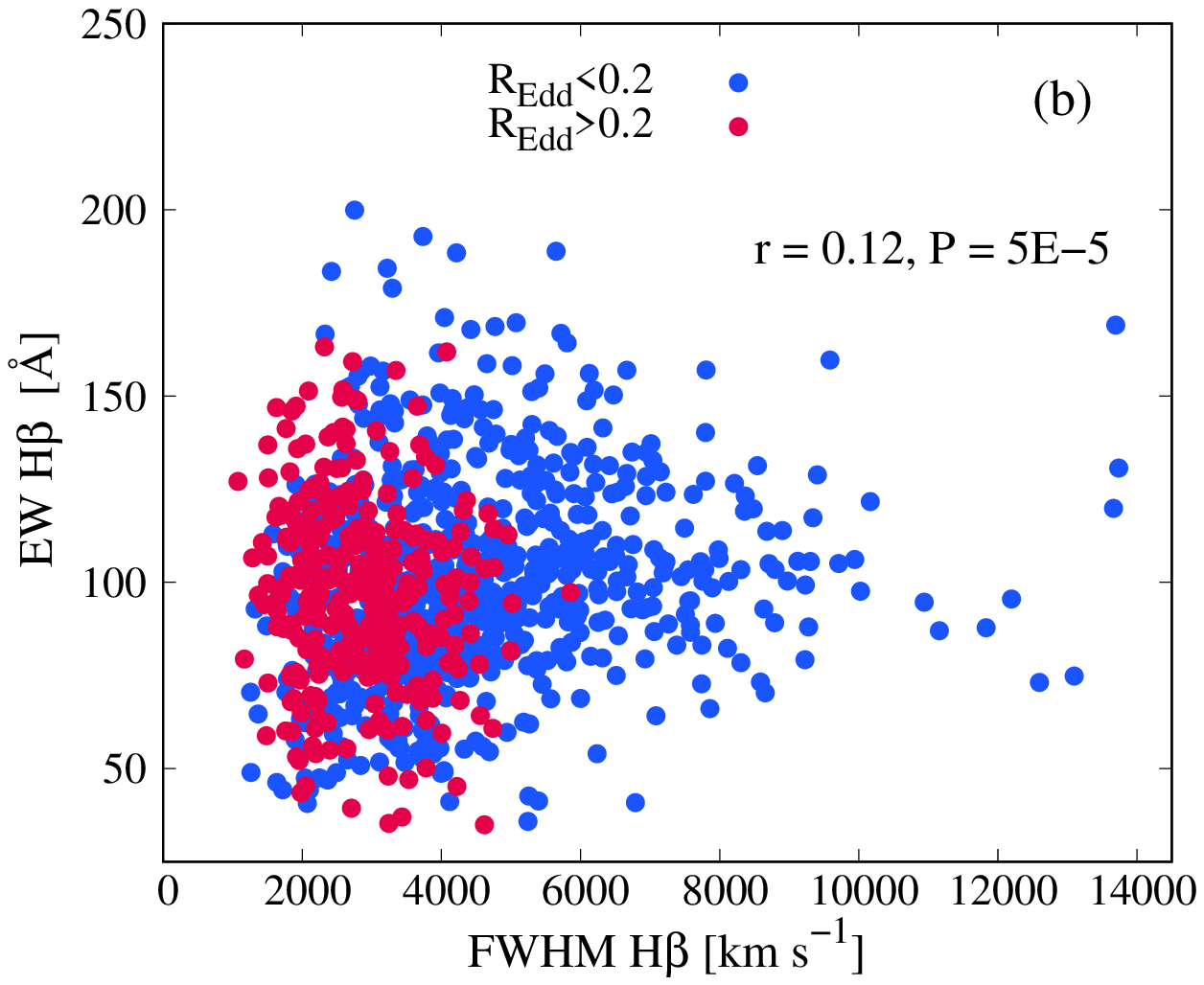}
   \caption{Relationship between EW Fe~II$_{tot}$ and FWHM Fe~II (a) and the same but for the EW H$\beta$ and FWHM H$\beta$ (b). The blue points denote objects with an Eddington ratio  lower than 0.2, while red points indicate sources with an Eddington ratio larger than 0.2. The Spearman coefficients of correlations (r) and P-values are given in graphs.}
    \label{fig02}
\end{figure}

Afterwards, we examined R$_{FeII}$ parameter, which is complex, since it represents the ratio of fluxes of two lines, Fe~II and H$\beta$. We checked what is relation between EWs of each of these two lines with their FWHMs (see Figure \ref{fig02}, Table \ref{T01}). We found that EW Fe~II$_{tot}$ versus FWHM Fe~II gives high anti-correlation (r = -0.53, P = 0), with very similar relationship between parameters as given in the quasar main sequence. On the other hand, EW H$\beta$ versus FWHM H$\beta$ shows no correlation. Since Eddington ratio (R$_{Edd}$) has been suggested as the driving mechanism of EV1 correlations, we investigated the role of R$_{Edd}$ in relationships of the EW Fe~II and EW H$\beta$ with their widths. Since the value R$_{Edd}\sim$ 0.2 is recognised in literature as possible critical value for which large changes occur in the accretion mode and structure of an AGN \citep{Wang2014,Ganci2019}, we plot dots in Figure \ref{fig02} with different colours for R$_{Edd}$ larger and smaller than 0.2. Results imply that objects with an R$_{Edd}$ larger than 0.2 are mainly objects with EW Fe~II$_{tot}$ > 150 \AA. We found no correlation between EW H$\beta$ and R$_{Edd}$, while EW Fe~II$_{tot}$ has moderate correlation with R$_{Edd}$ (r = 0.34, P = 0). As expected, the widths of both lines are in strong anti-correlation width R$_{Edd}$. The coefficient of correlation between R$_{Edd}$ and FWHM Fe~II is r = -0.51, P = 0, while for FWHM H$\beta$ it is r = -0.60, P = 0.

 We examined possible effects of mathematical coupling on the EW Fe~II-FWHM Fe~II anti-correlation, since both quantities  refer to the same line. Assuming a Gaussian shape for the Fe~II lines, these  two parameters are related as EW\,Fe\,II = F$_{FeII}$/F$_{cont}\sim$ (A$_{FeII}$/F$_{cont}) \cdot$ FWHM Fe~II, where F$_{FeII}$ and A$_{FeII}$ are the flux and amplitude of the Fe~II line, while F$_{cont}$ is the flux of the continuum under the Fe~II line. This implies that, if the ratio A$_{FeII}$/F$_{cont}$ remains approximately constant across the sample, EW Fe~II and FWHM Fe~II would exhibit a positive linear relationship due to the mathematical coupling between these two quantities. As seen in the data, a weak positive trend is observed for EW H$\beta$ versus FWHM H$\beta$, whereas EW Fe~II versus FWHM Fe~II relationship shows a strong negative, non-linear dependency, suggesting a physical origin (see Figure \ref{fig02}). This analysis implies that the simplified correlation underlying the quasar main sequence, which contains fewer parameters but represents the same physical background, is actually EW Fe~II versus FWHM Fe~II. This anti-correlation shows that as Eddington ratio increases, the Fe~II lines become narrower and the EW Fe~II grows. In further analysis, we focus to this anti-correlation  in order to analyse it in more details.

 \subsection{Inconsistent Fe~II lines in the contest of the quasar main sequence}\label{Sec3.2}

We investigated the role of different groups of Fe~II lines in the EW Fe~II versus FWHM Fe~II anti-correlation, and consequently in the quasar main sequence. First, we focused to the Fe~II$_{incons}$ lines, due to their specific behaviour in the spectra. It is found that for objects with low Eddington ratio and broad Fe~II lines,  Fe~II$_{incons}$ are weak or absent, as it is expected following their low transition probabilities. 
However, as Eddington ratio increases and line widths decrease, the intensity of these lines grows relative to Fe~II$_{cons}$ lines, and they become surprisingly strong, in some cases exceeding expected values by up to two orders of magnitude \citep{Kovacevic2025}.
 
\begin{figure}
	\centering
     \includegraphics[width=65mm, angle=270]{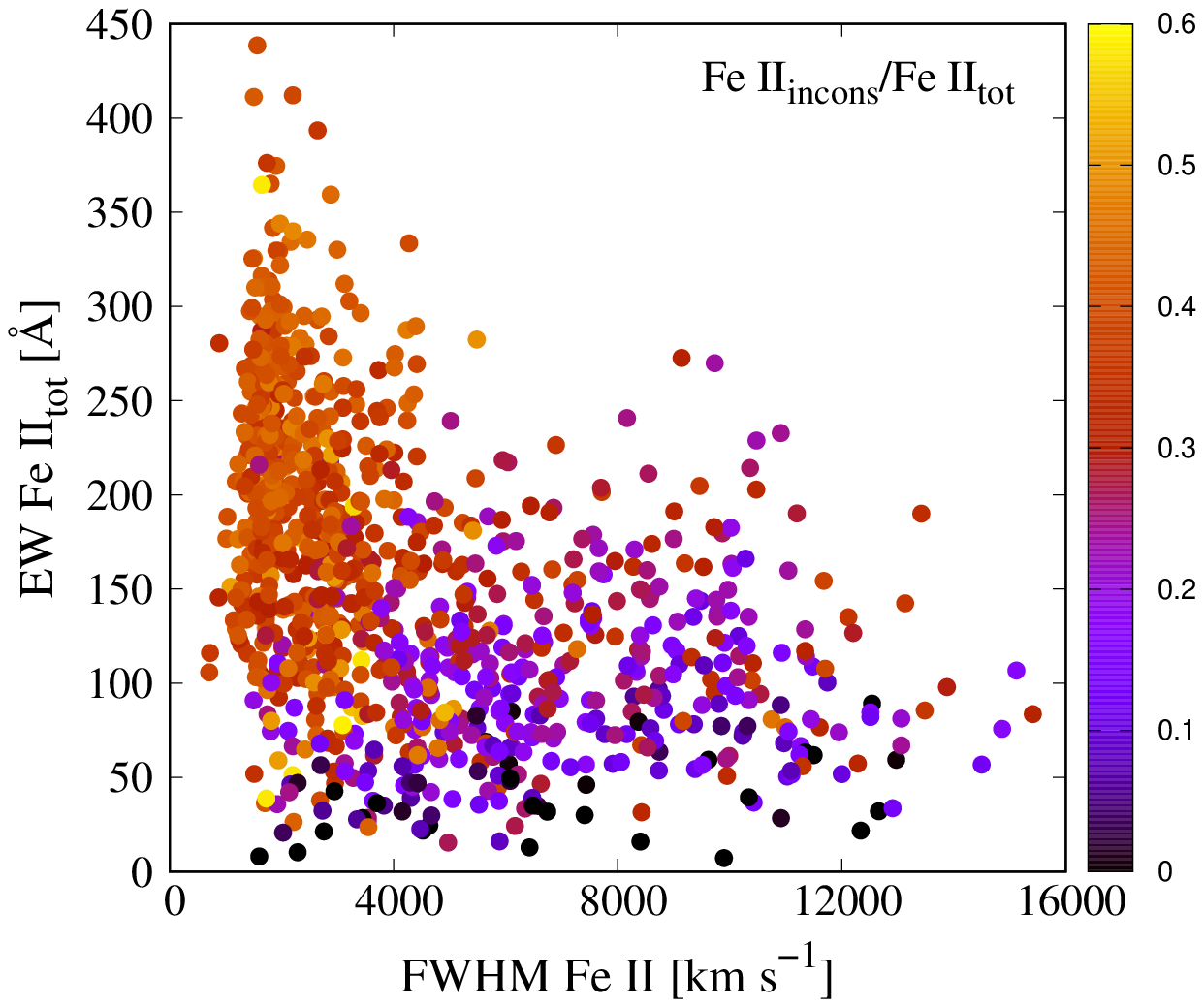}
    \caption{Contribution of the Fe~II$_{incons}$ lines to the total Fe~II flux in the 4000-5600 \AA \ range. Colours indicate the variation of the
    Fe~II$_{incons}$/Fe~II$_{tot}$ ratio across the EW Fe~II$_{tot}$--FWHM Fe~II parameter space.}
    
     \label{fig03}
\end{figure}

In Figure \ref{fig03} we plotted EW Fe~II$_{tot}$ versus FWHM Fe~II anti-correlation, coloured by contribution of Fe~II$_{incons}$ in the total Fe~II flux in the 4000-5600 \AA \ range. It could be seen that percentage of their contribution in total Fe~II flux grows for the objects with large EW Fe~II$_{tot}$, which have narrower Fe~II lines.  For objects with EW Fe~II$_{tot}$ > 150 \AA \ and FWHM Fe~II < 5000 km s$^{-1}$, the contribution of the  Fe~II$_{incons}$ makes up from 30\% to 50\% of the total Fe~II flux (see red and orange dots in Figure \ref{fig03}). On the other hand, for objects with smaller EW Fe~II$_{tot}$ and generally broader widths, the contribution of the Fe~II$_{incons}$ is mainly up to 20\% (blue and black dots in Figure \ref{fig03}).

\begin{figure}
	\centering
	\includegraphics[width=60mm, angle=270]{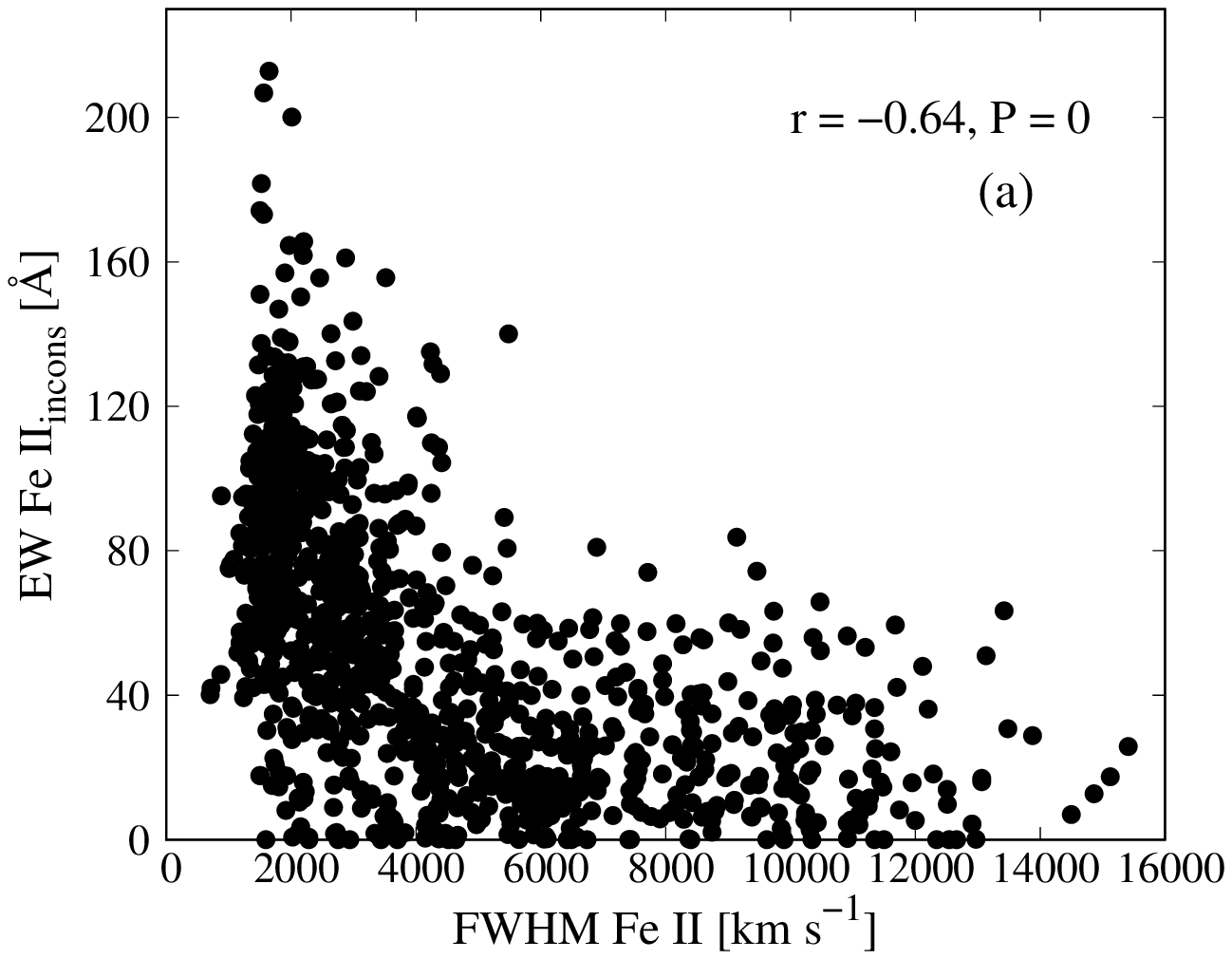}
	\includegraphics[width=60mm, angle=270]{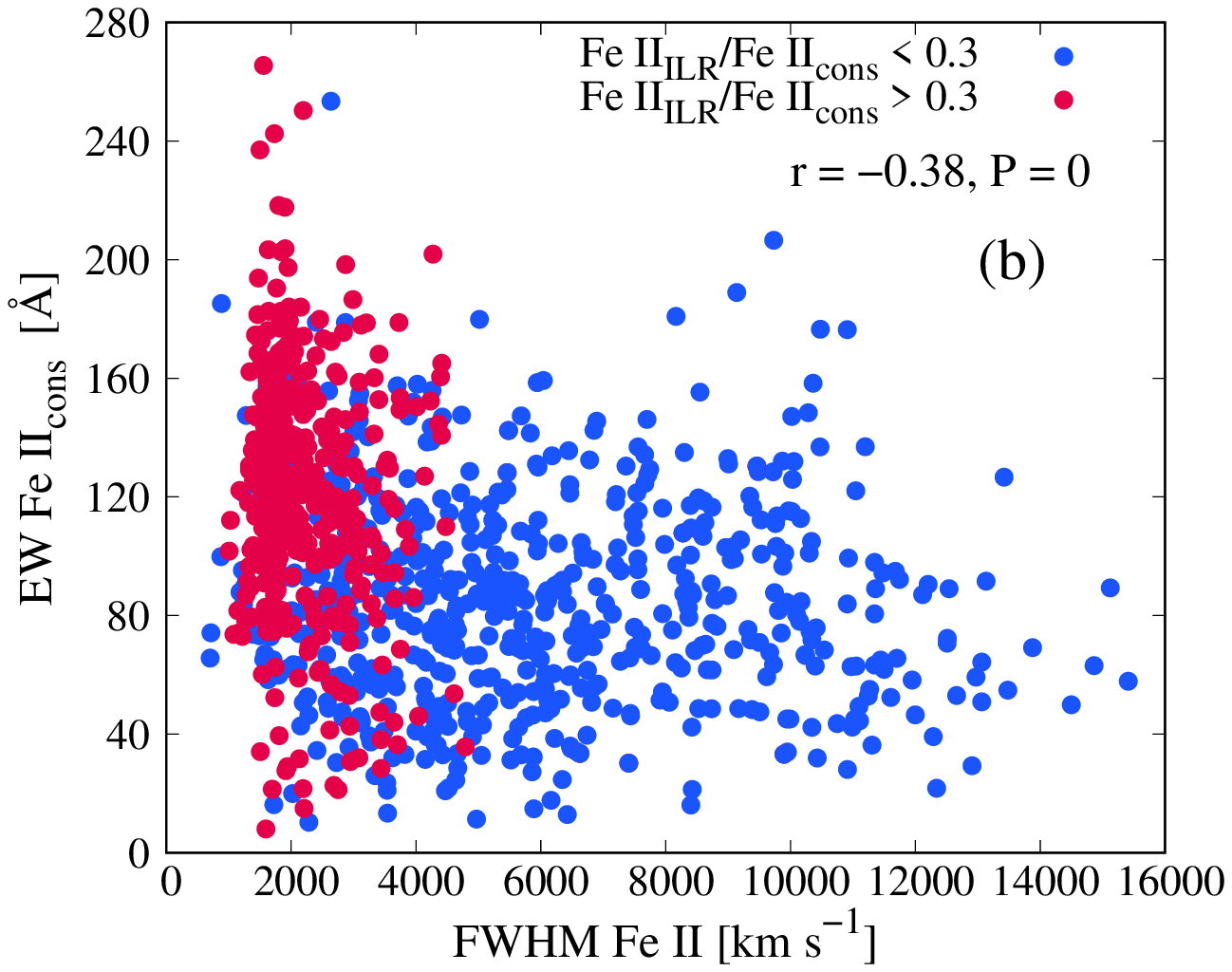}
 \caption{Correlation between the FWHM of Fe~II and the EW of Fe~II$_{incons}$ (a) and the EW of Fe~II$_{cons}$ (b). In panel (b), blue points denote objects in which the ILR component contributes less than 30\% of the Fe~II$_{cons}$ flux, while red points mark sources with an ILR contribution exceeding 30\% of the Fe~II$_{cons}$ flux.}
    \label{fig04}
\end{figure}

  These results implies that inconsistent Fe~II lines have important role in the growth of the total Fe~II flux for narrower Fe~II lines. Therefore we investigated separately the correlations of consistent and inconsistent Fe~II lines with the Fe~II width. We found that EW Fe~II$_{incons}$ lines are in significantly stronger anti-correlation with Fe~II width, than EW Fe~II$_{cons}$ (see Figure \ref{fig04}). By using only the EW of Fe~II$_{incons}$ lines in 4000-5600 \AA \ range, we found that the anti-correlation with FWHM Fe~II  becomes stronger than for total Fe~II, with coefficient of correlation r = -0.64, P = 0 (see Figure \ref{fig04} a). On the other hand, when using only the EW of consistent Fe~II lines, this anti-correlation is lower than for total Fe~II, and it becomes r = -0.38, P = 0 (see Figure \ref{fig04} b). The EW Fe~II$_{incons}$ are also in stronger anti-correlation with FWHM H$\beta$ than EW Fe~II$_{cons}$, and in significantly stronger correlation with R$_{Edd}$ (see Table \ref{T01}).

In order to additionally analyse and quantify the changes of different spectral parameters for spectra with narrower Fe~II,  we calculated the mean values for various spectral parameters in different Fe~II width ranges (see Table \ref{T03}). In the last column of the table we gave the 
ratios  between the mean parameter values for subsets with the broadest and with the narrowest Fe~II widths (FWHM Fe~II > 6000 km s$^{-1}$ and FWHM Fe~II < 2000 km s$^{-1}$, respectively). We found that all mean parameters related to the Fe~II flux increase rapidly for subsets with narrower Fe~II, while the mean EW H$\beta$ shows a slight decrease. The mean value of R$_{FeII}$ increases by a factor of 3.5, EW Fe~II$_{tot}$ by a factor of 2, EW Fe~II$_{cons}$ increases 1.5 times, while the mean value of the EW Fe~II$_{incons}$ shows the largest increase, by a factor of 3.6. The mean contribution of Fe~II$_{incons}$ in total Fe~II flux is doubled in the subset with narrowest Fe~II comparing the subset with the broadest Fe~II lines. The Fe~II$_{incons}$ in average make for 40\% of total Fe~II flux for subset with FWHM Fe~II < 2000 km s$^{-1}$.  The observed trends remain significant when the estimated 1$\sigma$ uncertainties in flux and EW of the various Fe~II components are taken into account. These results imply that although both consistent and inconsistent Fe~II lines grow for narrower spectra and higher R$_{Edd}$, the inconsistent Fe~II lines exhibit a more pronounced growth than the consistent ones.
  
 \subsection{Consistent Fe~II lines: Growth of the ILR contribution}\label{Sec3.3} 
 
 In our Fe~II model, the consistent Fe~II lines with estimated FWHM smaller than 5000 km s$^{-1}$, are fitted with two-components, ILR and VBLR, while for broader spectra we assume that Fe~II lines have only VBLR component. We tested the role of the Fe~II$_{ILR}$ and Fe~II$_{VBLR}$ components in growth of the Fe~II$_{cons}$,  in context of the quasar main sequence. In Figure \ref{fig04} b, the objects are  coloured in different colours according to the Fe~II$_{ILR}$ flux contribution in Fe~II$_{cons}$ flux. We found that objects with a Fe~II$_{ILR}$ flux contribution larger than 30\% dominate within the objects with a narrower FWHM Fe~II and at the same time with a large EW Fe~II$_{cons}$. 

The growth of the EW Fe~II$_{ILR}$ for objects with smaller Fe~II widths could also be seen in Table \ref{T03}, for subsets where Fe~II is modelled with  both components (FWHM Fe~II < 5000 km s$^{-1}$). The mean value of EW Fe~II$_{ILR}$ in the subset with FWHM Fe~II < 2000 km s$^{-1}$ is higher by a factor of 1.44 than in the subset with Fe~II line widths within the range 2000-4000 km s$^{-1}$.
Contrary of the Fe~II ILR component, there are no significant changes in the mean value of the EW Fe~II$_{VBLR}$ for subsets with different Fe~II widths. It seems that growth of the  EW Fe~II$_{cons}$ for objects with narrower Fe~II lines is dominantly caused by growth of the ILR components, while EWs of VBLR components in average remains the same. The Fe~II ILR components in average make for 38\% of  Fe~II$_{cons}$ flux for subset with FWHM Fe~II < 2000 km s$^{-1}$, while for subsets with broader Fe~II lines their contribution decreases.

\begin{figure}
	\centering
	\includegraphics[width=85mm]{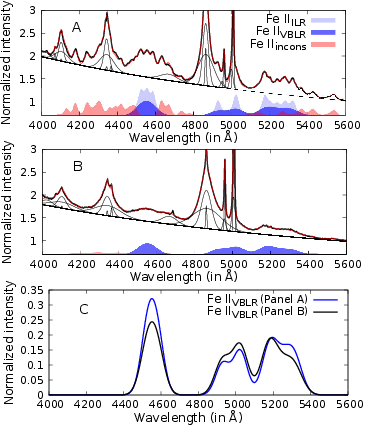}
 \caption{Decomposition of Fe~II lines in two mean spectra. Panel A: Mean spectrum obtained from the subset with EW Fe~II$_{tot}$ > 150 \AA \ and FWHM Fe~II < 5000 km s$^{-1}$. Panel B: Mean spectrum obtained from the subset with FWHM Fe~II > 5000 km s$^{-1}$. 
 The consistent Fe~II lines are shaded with the blue colour, where the sum of the ILR components is shaded with the light-blue, and the sum of the VBLR components with the dark-blue. The  inconsistent Fe~II lines are shaded with the red colour. In both panels, the best fit is denoted with the red line. The Fe~II$_{VBLR}$ components obtained from the best fit of the mean spectra are compared in Panel C. }
    \label{fig04_1}
\end{figure}

We additionally examined the role of the Fe~II ILR components in growth of the total Fe~II emission by performing one test. Following the distribution of parameters in EW Fe~II$_{tot}$ versus FWHM Fe~II plane (see Fig. \ref{fig03}), we selected two subsets with different properties, belonging to the two ends of this distribution. The first subset are the objects with the strongest Fe~II lines, selected to have EW Fe~II$_{tot}$ > 150 \AA \ and FWHM Fe~II < 5000 km s$^{-1}$ (418 objects). The second subset are the spectra with weak and very broad Fe~II lines, selected to have FWHM Fe~II > 5000 km s$^{-1}$ (338 objects). We re-binned all spectra, and normalised them to have the same continuum intensity at 5630 \AA, which is the only one adopted continuum window in fitting procedure. Afterwards, we found the mean spectrum for each subset, and we fit these two mean spectra with our model. In narrower mean spectrum, the Fe~II$_{cons}$ lines are fitted with the two-component model, while in the broad mean spectrum they are fitted with single-component model, assuming that they have only the Fe~II VBLR component. The best fits of the two mean spectra are shown in Fig. \ref{fig04_1}, where Fe~II$_{incons}$, Fe~II$_{ILR}$ and Fe~II$_{VBLR}$ components are shaded with different colours. We found that inconsistent Fe~II lines are strong in the mean spectrum with strong and narrow Fe~II (Fig. \ref{fig04_1}, Panel A), while they are negligible in the broad line mean spectrum (Fig. \ref{fig04_1}, Panel B), where only Fe~II$_{VBLR}$ contributes to the total Fe~II emission. 

However, most interestingly, the Fe~II VBLR component of the mean spectrum made from the subset with EW Fe~II$_{tot}$ > 150 \AA \ and FWHM Fe~II < 5000 km s$^{-1}$,  is very similar to the Fe~II VBLR component of the mean spectrum from the  FWHM Fe~II > 5000 km s$^{-1}$  subset (see Fig. \ref{fig04_1}, Panel C). The width of the Fe~II VBLR component in narrow mean spectrum is $\sim$ 5400 km s$^{-1}$, and in broad one is $\sim$ 5900 km s$^{-1}$, while both EWs are $\sim$ 100 \AA. These results additionally illustrate that the very strong EW Fe~II in some objects is dominantly caused by the enhanced emission of the EW Fe~II$_{incons}$ and EW Fe~II$_{ILR}$, while the EW Fe~II$_{VBLR}$  on average remains the same in the entire sample.

 \begin{table} [!t]
\begin{center}
\caption{Correlations between different properties of Fe~II and H$\beta$ lines and some physical parameters.  
\label{T01}}
\footnotesize
\begin{tabular}{|c |c |c |c |c |}

\hline

 & &FWHM Fe~II & FWHM H$\beta$  &  logR$_{Edd}$ \\
\hline
 \hline

\multirow{2}{*}{R$_{FeII}$} &r  & -0.57  & -0.34 & 0.33 \\
    &P  &  0  & 0 & 0 \\

\hline

\multirow{2}{*}{Fe~II$_{tot}$/H$\beta$} &r  & -0.56  & -0.34 & 0.33 \\
    &P  &  0  & 0 & 0 \\
\hline

    \multirow{2}{*}{ FWHM Fe~II } &r  & 1  &0.66 & -0.51  \\
    &P  &  0 &0  & 0 \\

\hline  

\multirow{2}{*}{ FWHM H$\beta$ } &r  & 0.66  &1 & -0.60 \\
    &P  &  0  & 0  & 0 \\
    
\hline 

\multirow{2}{*}{ EW H$\beta$} &r  & 0.32  &0.12 & -0.12 \\
    &P  &  0 & 5E-5  & 1E-4 \\

\hline

\multirow{2}{*}{ EW Fe~II$_{tot}$} &r  & -0.53  & -0.34 & 0.34 \\
    &P  &  0  &0 & 0 \\

\hline

\multirow{2}{*}{ EW Fe~II$_{cons}$} &r  & -0.38  & -0.22 & 0.27 \\
    &P  &  0  & 3E-13 & 0 \\

\hline

\multirow{2}{*}{ EW Fe~II$_{incons}$} &r  & -0.64  & -0.43 & 0.40 \\
    &P  & 0 & 0  & 0 \\
\hline

\hline

\end{tabular}
\tablefoot{Table contains Spearman coefficients of correlation (r) and P-values. }

\end{center}
\end{table}

 \begin{table*}

\begin{center}
\caption{Mean values of the measured parameters for  different FWHM Fe~II ranges.
\label{T03}}
\smaller[1]

\begin{tabular}{| c |c |c |c |c |c |}
\hline

     (1) &               (2)         &     (3)       & (4)  &   (5) & (6)\\ [1mm]
     \hline
\multirow{2}{*}  {\rm Width range}    &               FWHM FeII>6000          &     4000<FWHM FeII<6000       & 2000<FWHM FeII<4000  &   FWHM FeII<2000 & \multirow{2}{*}  {$\frac{(5)}{(2)}$}\\

 &      [255]               &  [175]          & [327]  &   [256] &  \\  [1mm]
\hline
\hline

\rule{0pt}{3ex} FWHM Fe~II                & 8969  $\pm$ 2201   &         4901 $\pm$ 603                 & 2827 $\pm$ 524                       &   1648 $\pm$ 234  & 0.18 \\[1mm]
 \hline
\rule{0pt}{3ex} FWHM H$\beta$                   &5778  $\pm$ 1965&  4023 $\pm$ 1311               & 3463 $\pm$ 1155                        & 2746 $\pm$ 930     &   0.47  \\  [1mm]
 \hline       
\rule{0pt}{3ex}  R$_{FeII}$  &  0.36  $\pm$ 0.19     &     0.41  $\pm$ 0.23     & 0.59  $\pm$ 0.29    &  0.91 $\pm$ 0.43  & 2.53 \\[1mm]
   \hline 
\rule{0pt}{3ex}  EW Fe~II$_{tot}$                &  108  $\pm$ 49                &  119 $\pm$ 61             & 167 $\pm$ 78                       & 215 $\pm$ 84  & 2.00   \\   [1mm]
   \hline       
 
\rule{0pt}{3ex} EW Fe~II$_{cons}$                &84  $\pm$ 34                        & 85 $\pm$ 37                    & 103 $\pm$ 47                         & 129 $\pm$ 49   & 1.54 \\  [1mm]
  \hline           

\rule{0pt}{3ex} EW FeII$_{incons}$                &24  $\pm$ 18                       & 34 $\pm$ 29                  & 65 $\pm$ 35                       & 87 $\pm$ 37   &   3.62  \\ [1mm] 
  \hline

\rule{0pt}{3ex} Fe~II$_{incons}$/Fe~II$_{tot}$                & 0.20  $\pm$ 0.11                       & 0.26 $\pm$ 0.12                       & 0.37 $\pm$ 0.10 & 0.40 $\pm$ 0.06  &  2.00  \\ [1mm]
 \hline        
 
\rule{0pt}{3ex} EW Fe~II$_{ILR}$                & --       & 7 $\pm$ 23                & 34 $\pm$ 26                            & 49 $\pm$ 24  &  --  \\[1mm]
  \hline

\rule{0pt}{3ex} EW Fe~II$_{VBLR}$                & 82  $\pm$ 35                 & 77 $\pm$ 34                    & 69 $\pm$ 37                      & 80 $\pm$ 32 & 0.98\\ [1mm]
  \hline

\rule{0pt}{3ex} Fe~II$_{ILR}$/Fe~II$_{cons}$                & --  & 0.06 $\pm$ 0.18 & 0.31 $\pm$ 0.21 &0.38 $\pm$ 0.13 & --  \\   [1mm]   
  \hline

\rule{0pt}{3ex} EW H$\beta$                     & 110 $\pm$ 26                 & 102 $\pm$ 25                    & 101 $\pm$ 29                          & 88 $\pm$ 36 &  0.80  \\[1mm]
  \hline

\rule{0pt}{3ex} R$_{Edd}$                 &   0.09  $\pm$ 0.08        &   0.13  $\pm$ 0.12              &  0.20  $\pm$ 0.16       &  0.30  $\pm$ 0.25   &    3.33             \\[1mm]

 \hline

\end{tabular}
\tablefoot{Table contains mean values $\pm$ standard deviation. The number of objects in each FWHM Fe~II range is given in brackets.  The FWHMs of lines are given in km s$^{-1}$, and EWs are given in \AA. Column (6) gives the ratios between the mean values obtained for the subsets with the narrowest and with the broadest Fe~II widths (columns 5 and 2).
}

\end{center}
\end{table*}

 \subsection{Fe~II lines: A signature of the AGN dichotomy}\label{Sec3.4} 
 
As previously noted in Section \ref{Sec1}, it has been observed that R$_{FeII}$ versus FWHM H$\beta$ anti-correlation can be utilised for classification purposes \citep{Sulentic2000a, Marziani2001}. \cite{Sulentic2000a} suggested that the specific shape of this anti-correlation, with two visually distinct parameter branches in the R$_{FeII}$ versus FWHM H$\beta$ plane, reveals a division into two AGN populations. These two populations are visually separated at FWHM H$\beta$ = 4000 km s$^{-1}$, where Population A has FWHM H$\beta$ < 4000 km s$^{-1}$ and Population B, FWHM H$\beta$ > 4000 km s$^{-1}$. The comparison of different spectral properties in radio, optical, UV and X-range among these two populations, indicated significant differences  among these two groups of objects \citep{Sulentic2011,Marziani2018}.
 
We found that  EW Fe~II versus FWHM Fe~II anti-correlation has higher coefficient of correlation comparing R$_{FeII}$ versus FWHM H$\beta$, implying that it is probably more fundamental. Therefore, we investigated whether it also could reflect the AGN dichotomy. 
 The results presented in Fig. \ref{fig02}a, Fig. \ref{fig03} and Fig. \ref{fig04}b reveal that the EW Fe~II versus FWHM Fe~II correlation space indicate two branches of the data. One branch is with FWHM Fe~II < 5000 km s$^{-1}$ and strong EW Fe~II (>150 \AA), which is distinguished with higher Eddington ratio, and higher contribution of the Fe~II$_{incons}$ and Fe~II$_{ILR}$ in  Fe~II$_{tot}$ flux. As it could be seen in Table \ref{T03}, the growth of the EW Fe~II$_{tot}$ in this group is mainly caused by the growth of EW Fe~II$_{incons}$ and EW Fe~II$_{ILR}$. The other branch in EW Fe~II versus FWHM Fe~II plane represents the objects with smaller R$_{Edd}$ and smaller contribution of the Fe~II$_{incons}$ and Fe~II$_{ILR}$ in total EW Fe~II.  In these objects, EW Fe~II$_{tot}$ mainly consists of EW Fe~II$_{VBLR}$, and it goes up to $\sim$ 150 \AA \ with average value of $\sim$ 100 \AA. This group mostly includes the objects with broad FWHM Fe~II ( > 5000 km s$^{-1}$),  weak EW Fe~II$_{tot}$ and low Eddington ratio. At the breaking point of these two branches of the data are objects with FWHM Fe~II < 5000 km s$^{-1}$ and weak EW Fe~II (<150 \AA), which represents a mixture of these two groups.

\begin{figure}
	\centering
	\includegraphics[width=60mm, angle=270]{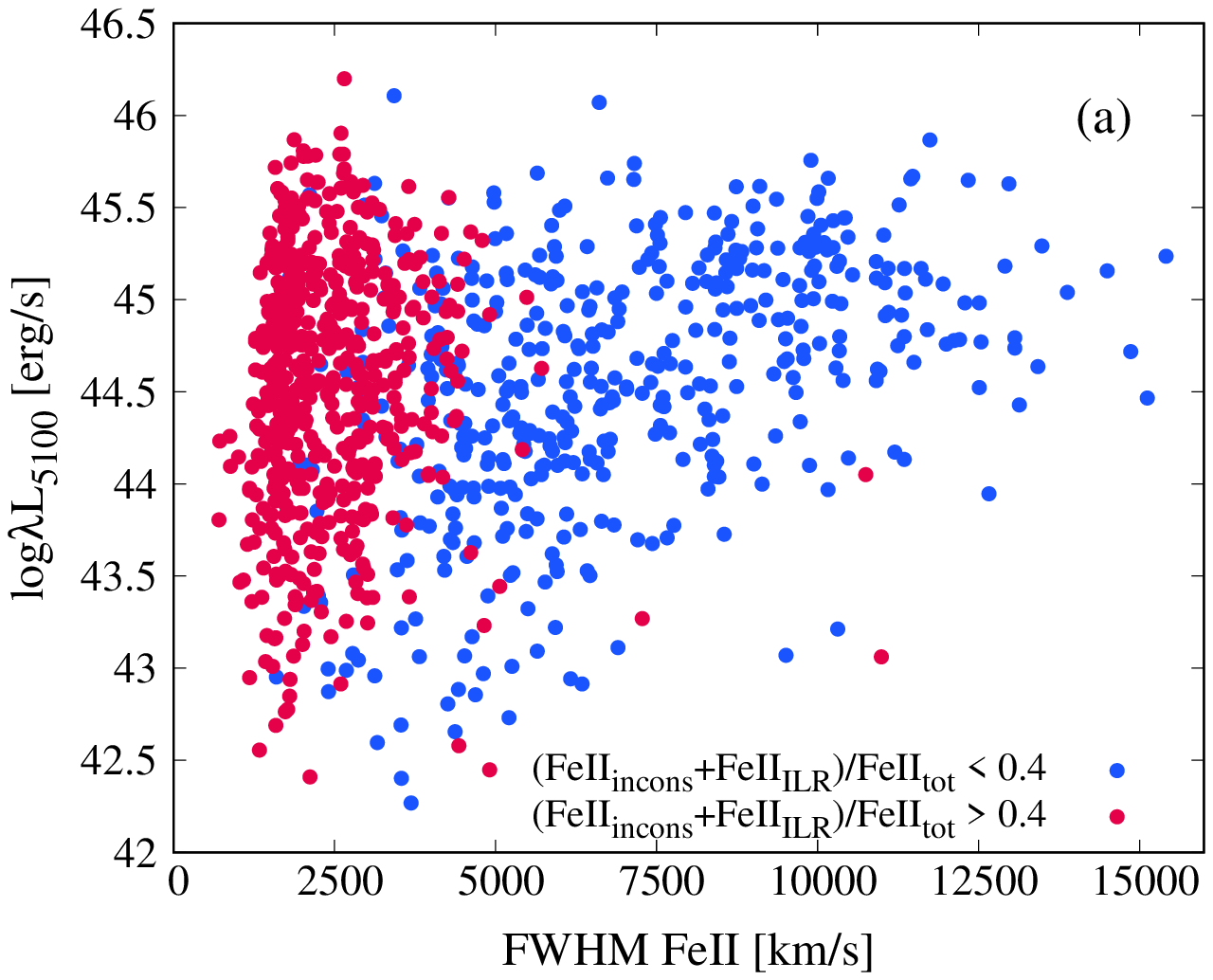}
	\includegraphics[width=60mm, angle = 270]{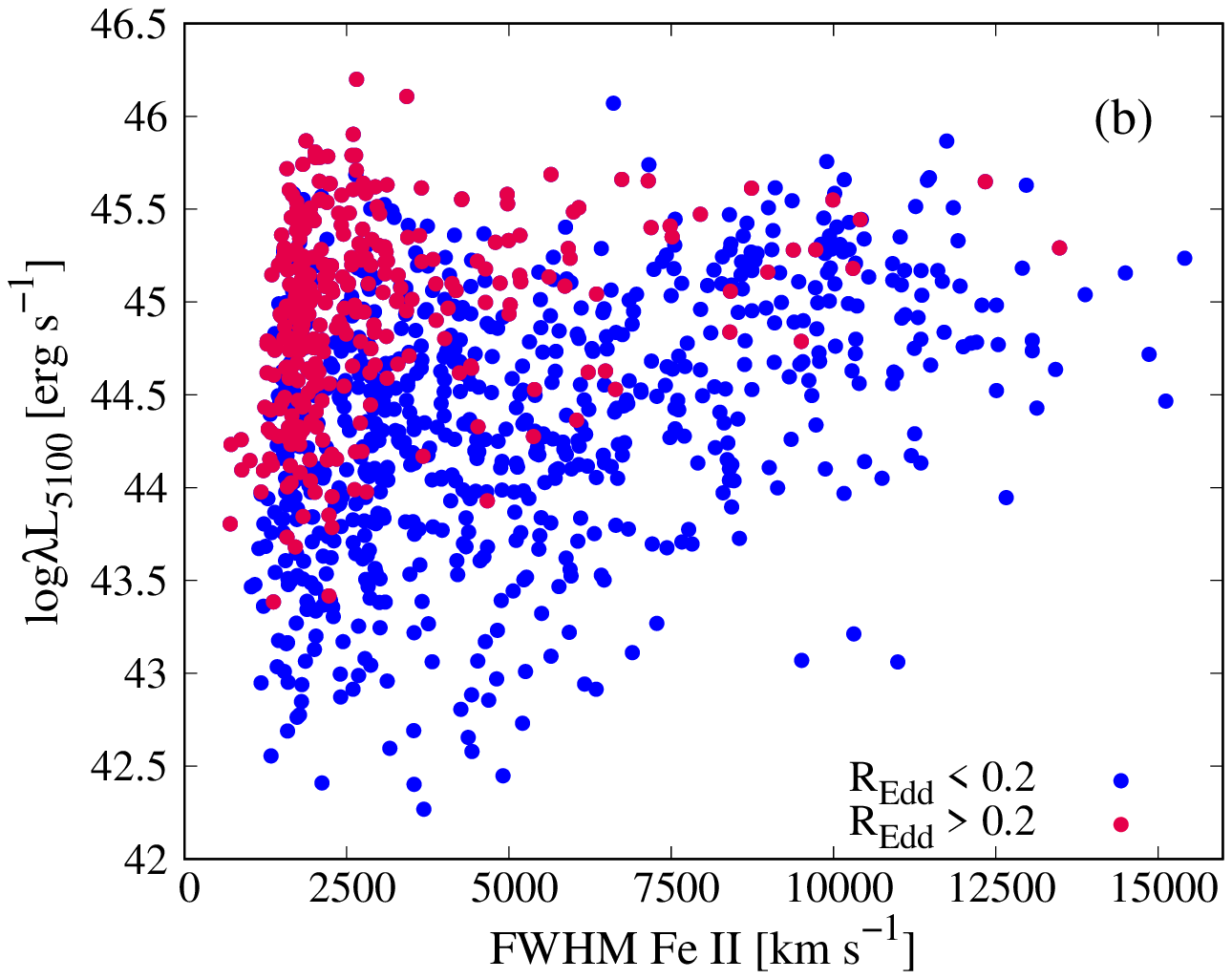}
 \caption{Relationship between log$\lambda L_{5100}$ and the width of the Fe~II lines. In panel (a), red points mark objects in which the combined contribution of Fe~II$_{incons}$ and Fe~II$_{ILR}$ exceeds 40 \% of the total Fe~II flux, whereas blue points denote sources in which this contribution falls below the same threshold. The distribution of objects with R$_{Edd}$ smaller and larger than 0.2 is shown in panel (b).}
    \label{fig05}
\end{figure} 

It seems that those two groups of objects are even more clearly separated in FWHM Fe~II versus log$\lambda L_{5100}$ parameter space (see Fig. \ref{fig05}), where they are positioned in two visually distinct branches with different dependencies between FWHM Fe~II and luminosity, forming a heart-like shape.  The objects with contribution of the sum of Fe~II$_{incons}$ and Fe~II$_{ILR}$ larger than $\sim$40\% in the total Fe~II flux are concentrated in the left branch and have steeper FWHM Fe~II versus log$\lambda L_{5100}$ relationship. As expected, these objects generally exhibit higher Eddington ratios than those distributed along the right branch of the diagram. Most sources with Eddington ratios exceeding the critical value of R$_{Edd}$ $\sim$ 0.2 are located on the left branch (see Figure \ref{fig05}b). 

 We investigated whether objects populating the two branches of the heart-shaped diagram differ in their spectral properties and global parameters, and compared these trends with those found for the Pop A/Pop B dichotomy. It has been found that Population A sources generally have lower black hole masses and enhanced outflow signatures relative to Population B sources \citep{Marziani2018, Shen2014}. Also, it has been suggested that orientation effects may contribute to the dispersion in FWHM H$\beta$ along the quasar main sequence at fixed R$_{FeII}$ \citep{Shen2014}. In contrast to the  Pop A/Pop B division, we found no difference  in the M$_{BH}$ distributions for objects in two branches of FWHM Fe~II versus log$\lambda L_{5100}$ diagram. We examined the possible presence of AGN-driven outflows by analysing the complex profile of the [O~III] 5007 \AA \ line, which is known to be a good indicator of outflow presence and strength \citep[see e.g.][]{Woo2016, Rakshit2018}. We used the velocity shift of the [O~III] 5007 \AA \ wing component to trace the outflow contribution. The results show that [O~III] wing components are more blueshifted in the left branch of the heart-shaped diagram indicating the outflow presence. The details of analysis and plot are given in Appendix \ref{B}, Figure \ref{B1}. \cite{Shen2014} used the ratio of the FWHM H$\beta$ and the second moment of the emission-line profile of the same line ($\sigma$ H$\beta$) as one of the indicators of the orientation effects. The variation of this ratio traces changes in the H$\beta$ profile, which is known to differ systematically between Population A and Population B sources \citep{Marziani2018}.
\cite{Shen2014} suppose that in the case of the disc-like BLR geometry, $\sigma$ H$\beta$ is less affected by inclination effects leading to a larger FWHM H$\beta$ /$\sigma$ H$\beta$ ratio as inclination increases. We computed this ratio for our sample and found that it follows the dichotomy of the heart-shaped diagram, where sources with lower FWHM H$\beta$ /$\sigma$ H$\beta$ values  are predominantly located in the left branch (see Appendix \ref{B}, Figure \ref{B2}), indicating smaller inclinations for these objects. However, these variations in H$\beta$ profile, which distinguish Pop A and Pop B sources, as well as  objects in two branches of the FWHM Fe~II versus log $\lambda L_{5100}$ diagram, may also arise from factors other than inclination, such as structural differences in the accretion disc or the BLR \citep{Collin2006}.

Additionally, we focus on the well known  anti-correlation between EW Fe~II and EW [O~III], which is part of EV1 correlations found by \cite{Boroson1992}, and we tried to analyse it from an another angle. In Figure \ref{fig08} we coloured two relationships, EW Fe~II versus FWHM Fe~II and log$\lambda L_{5100}$ versus FWHM Fe~II, with different colours according to the value of the [O~III]$_{5007 \AA}$/ H$\beta_{NLR}$ ratio. This ratio could be considered as an indicator of the shape of the ionisation continuum \citep{Baldwin1981} in the region where narrow lines arise. We found that objects with narrow and strong EW Fe~II, which are also the objects in the left branch of the log$\lambda L_{5100}$ versus FWHM Fe~II diagram, mostly have smaller [O~III]/H$\beta_{NLR}$ ratio, comparing the rest of the objects (see Figure \ref{fig08}). For the narrowest and strongest Fe~II, this ratio becomes even log([O~III]/H$\beta_{NLR}$)<0.5. Some of these objects correspond to NLSy1 galaxies, which are defined as  having [O~III]$_{5007}$/H$\beta_{NLR}$)<3 and FWHM H$\beta$<2000 km s$^{-1}$ \citep{Osterbrock1985, Goodrich1989}. In our sample, there are 22 objects with these properties and they are located in the left branch of the diagram, between the sources with the narrowest Fe~II line widths  ($\sim$1500 km s$^{-1}$) and with log$\lambda L_{5100}$ up to 44.5 erg s$^{-1}$.

Finally, we compared the dichotomy seen in the heart-shaped diagram with classical division into Pop A/Pop B (see Fig. \ref{fig08_1}). We find that the left and right branches of the diagram approximately follow the Pop A/Pop B classification. Pop A sources are mostly located in the left branch, Pop B in the right one, while in the  FWHM Fe~II range between 4000-6000 km s$^{-1}$, there is a mixture of  Pop A and Pop B sources.

\section{Discussion}\label{Sec4}

\subsection{Dissecting the R$_{FeII}$ parameter}\label{Sec4.1}

The principal component analysis performed by \cite{Boroson1992} revealed that the parameter  R$_{FeII}$ is governed by Eddington ratio. 
Since then, the parameter R$_{FeII}$ has been widely discussed in the literature in the context of its significant relationship with Eddington ratio and accretion rate, although the nature of this relationship is not yet understood \citep[see e.g.][]{Sulentic2000a, Boroson2002, Shen2014, Yu2020,  Marziani2025}. It is proposed that  R$_{FeII}$ could be used for improving the accuracy of the M$_{BH}$ estimation  \citep{Du2019, Yu2020}. Namely, it is found that objects with high Eddington ratio deviate from the empirical R$_{BLR}$–L$_{5100}$ relationship found in \cite{Kaspi2000} and used for M$_{BH}$ estimation \cite[see][]{Peterson2004}. Since  R$_{FeII}$ was found to correlate with that deviation,  it is proposed to use the parameter R$_{FeII}$ to improve the R$_{BLR}$–L$_{5100}$ relationship \citep{Du2019, Yu2020}. On the other hand, \cite{Marziani2014} proposed the parameter  R$_{FeII}$  as an indicator of objects with highest value of Eddington ratio, defined as extreme Population A sources, with R$_{FeII}$ > 1. These objects are characterised by a number of specific properties that distinguish them from other objects in the AGN population \citep[see][]{Marziani2025}.

 \begin{figure}
	\centering
	\includegraphics[width=60mm, angle = 270]{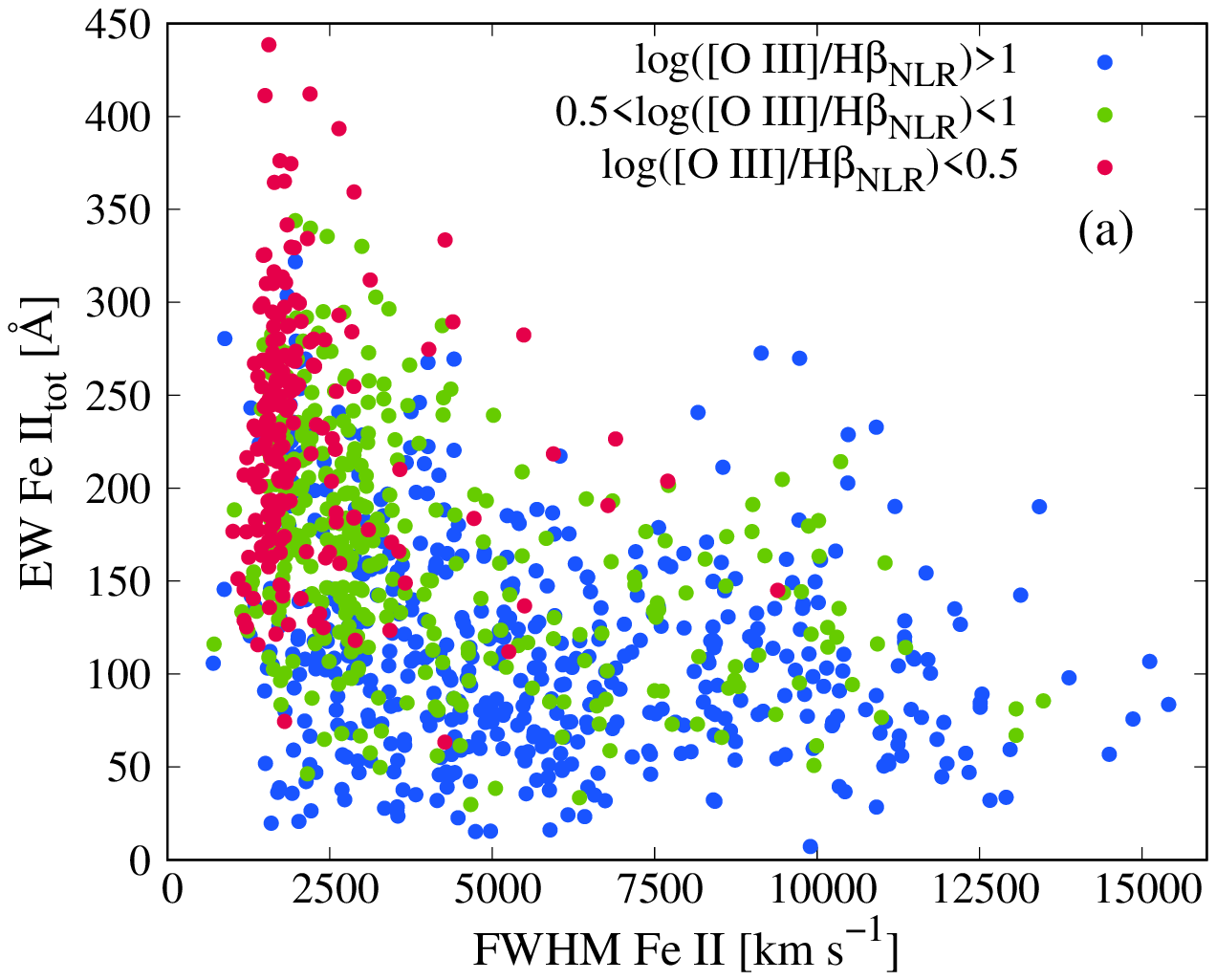}
	\includegraphics[width=60mm, angle = 270]{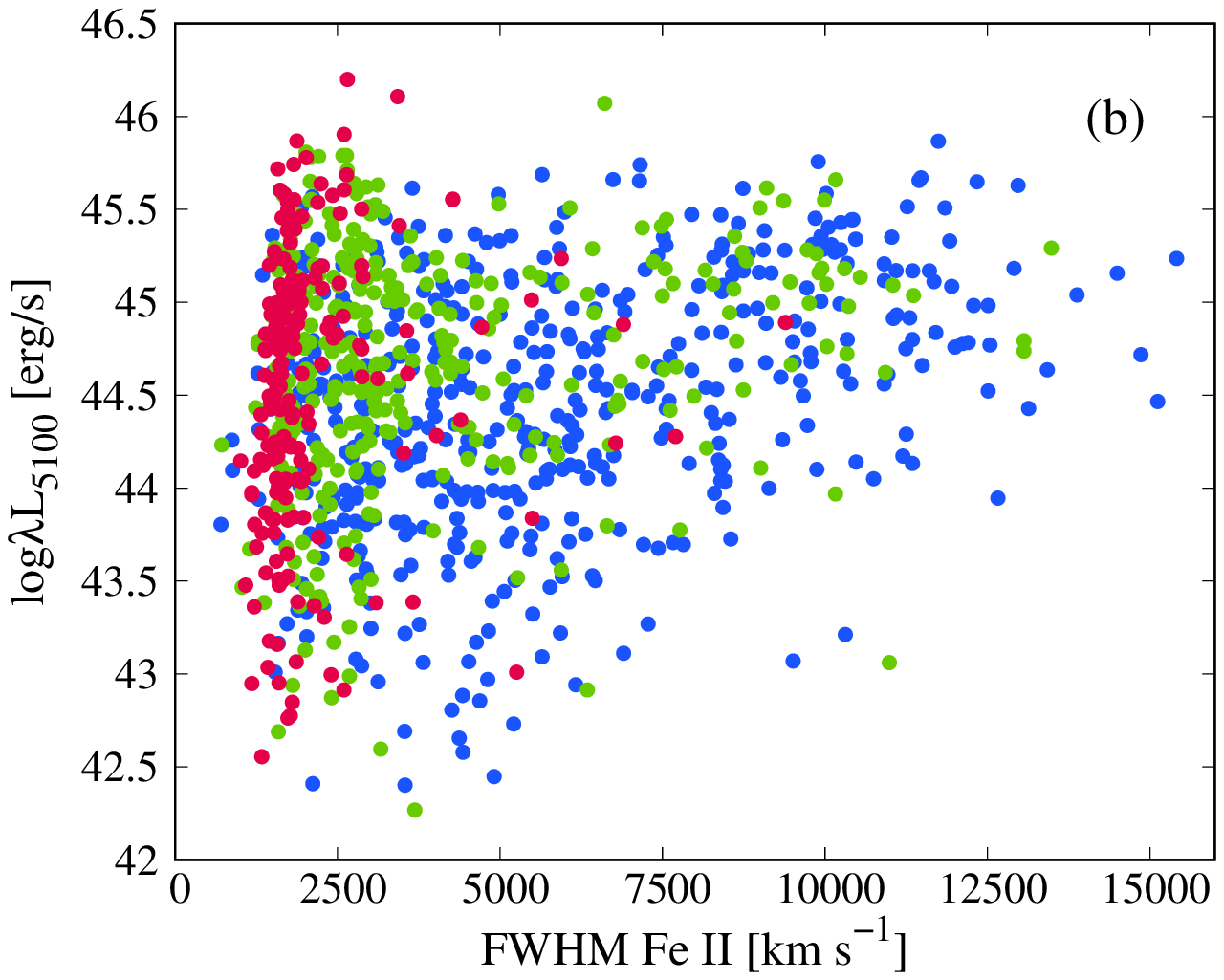}
 \caption{Variation of the [O~III]/H$\beta_{NLR}$ ratio across the FWHM Fe~II--EW Fe~II$_{tot}$ (a) and FWHM Fe~II--log$\lambda L_{5100}$ (b) parameter space. The different colours of the points indicate the strength of the log[O~III]/H$\beta_{NLR}$ ratio. }
    \label{fig08}
\end{figure} 

In order to better understand the nature of the relationship between R$_{FeII}$ and R$_{Edd}$, and the role of the Eddington ratio in quasar main sequence correlation, we applied the flexible and complex Fe~II model, which enabled us  to precisely measure the width of the Fe~II lines and to follow the changes among the relative intensities of different Fe~II line groups along the quasar main sequence.   
Our results imply that a simpler and more fundamental relationship underlying the anti-correlation of R$_{FeII}$ versus FWHM H$\beta$ may be the EW Fe~II versus FWHM Fe~II, where EW Fe~II becomes even twice as strong in AGNs with narrow Fe~II lines and high Eddington ratio, compared to those with low Eddington ratio.  Since there is no significant correlation between EW H$\beta$ and Eddington ratio, this implies that the EW Fe~II parameter is itself an indicator of the R$_{Edd}$ growth.  This raises the question of why EW Fe~II may represent a more direct physical link to R$_{Edd}$ than  EW H$\beta$.  This may be related to the exceptionally complex atomic structure of Fe~II,  characterised by a very large number of closely spaced energy levels and, consequently, numerous possible transitions. This property makes Fe~II distinct from other atomic species and enables a wide variety of atomic processes  that can be triggered under specific astrophysical conditions. At the same time, an increase in R$_{Edd}$ may significantly modify the physical conditions and structure of the BLR. Such changes could activate additional atomic processes in Fe~II or enhance the efficiency of existing ones, leading to additional iron emission.

Based on the shape of the quasar main sequence, it could be seen that a small FWHM H$\beta$ (and thus a small FWHM Fe~II) appears to be a necessary, but not sufficient condition for a high R$_{FeII}$ value.  In the FWHM Fe~II versus EW Fe~II parameter space, the objects with an EW Fe~II less than $\sim$150 \AA \  are uniformly present for all widths of the Fe~II lines, while the strong Fe~II emitters (EW Fe~II larger than $\sim$150 \AA) appear only for objects with narrower lines (FWHM Fe~II < 5000 km s$^{-1}$), followed with a high Eddington ratio.

\begin{figure}
	\centering
	\includegraphics[width=60mm, angle = 270]{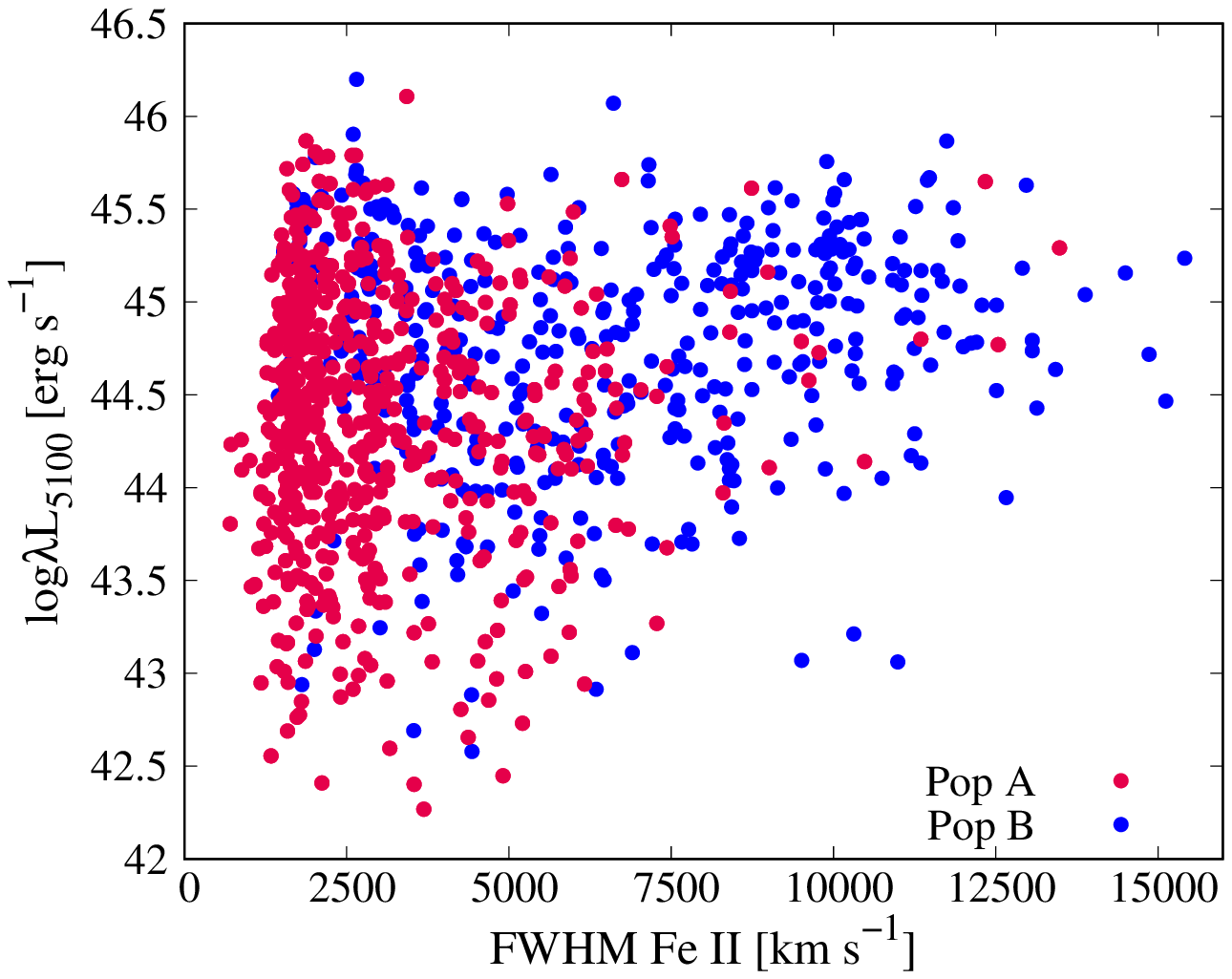}
 \caption{Distribution of the Pop A and Pop B objects in the FWHM Fe~II versus log $\lambda L_{5100}$ plane.}
    \label{fig08_1}
\end{figure}

\begin{figure}
	\centering
	\includegraphics[width=85mm]{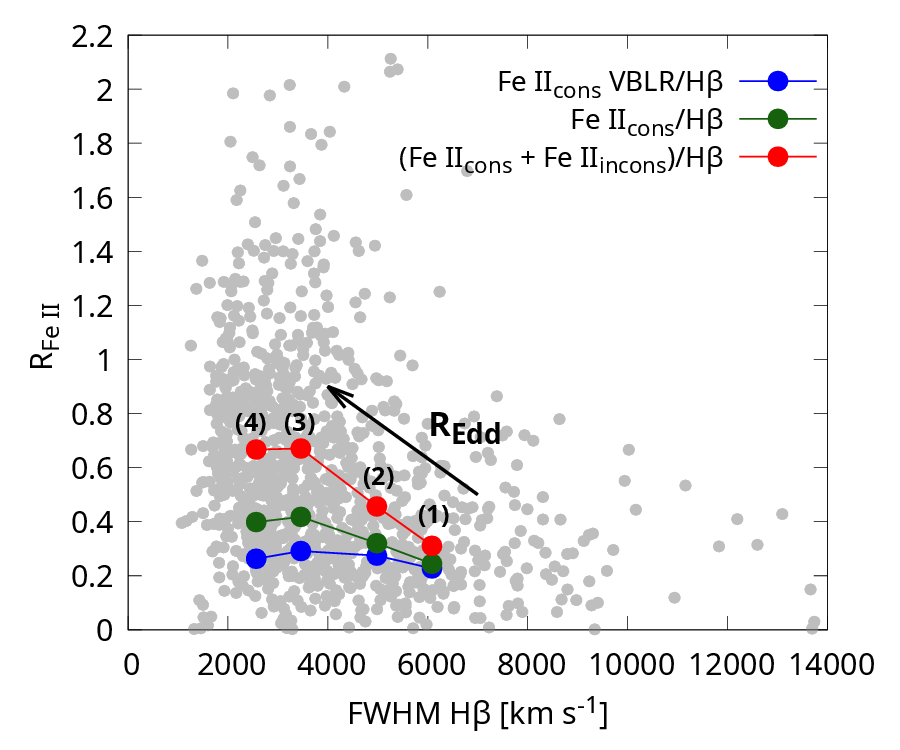}
 \caption{Growth of different Fe~II components in the quasar main sequence. The sample is divided into four subsets according to R$_{Edd}$: (1) R$_{Edd}$ < 0.03 (71 objects),  (2) 0.03 < R$_{Edd}$ < 0.1 (325 objects), (3) 0.1 < R$_{Edd}$ < 0.3 (471 objects), and (4) R$_{Edd}$ > 0.3 (154 objects). For each subset, we found the mean ratio of different Fe~II components in 4434-4684 \AA \ range and broad H$\beta$ (marked with different colours), as well as the mean of the  FWHM H$\beta$. The mean values of each subset are indicated by the corresponding numbers (1-4) on the top and plotted over the quasar main sequence. }
    \label{fig07}
\end{figure}

We separately analysed the behaviour of the EW Fe~II$_{incons}$ and EW Fe~II$_{cons}$ along the main sequence, and additionally, within the EW Fe~II$_{cons}$, we analysed the changes in their line components,  EW Fe~II$_{ILR}$  and EW Fe~II$_{VBLR}$. We found that the EW Fe~II$_{VBLR}$  components remain on average unchanged along the quasar main sequence, while the EW Fe~II$_{incons}$  and EW Fe~II$_{ILR}$  increase for smaller line widths (see  Figure \ref{fig04_1} and Table \ref{T03}), and high Eddington ratio.  This is additionally demonstrated in Figure \ref{fig07}, where the sample is divided into different subsets according to the R$_{Edd}$, from subset with the smallest R$_{Edd}$ (assigned as (1) on the plot) to the subset with the largest R$_{Edd}$ (assigned as (4) on the plot). The blue dots represent the mean values of the Fe~II$_{VBLR}$/H$\beta$ ratio, the green dots are the mean values of Fe~II$_{cons}$/H$\beta$ (where both components are included, ILR+VBLR). Finally, when we included Fe~II$_{incons}$, we obtained the red dots, which represent the mean values of R$_{FeII}$, i.e. the ratio of total Fe~II in the range 4434-4684 \AA \  (Fe~II$_{cons}$ + Fe~II$_{incons}$) divided by H$\beta$. It could be seen that the mean values of the EW Fe~II$_{VBLR}$  do not change much for different R$_{Edd}$ (blue dots), while EW Fe~II$_{ILR}$  slightly increases for higher R$_{Edd}$ (green dots).  However, the largest increase is in the EW Fe~II$_{incons}$ lines (red dots), which cause the mean value of the R$_{FeII}$ to almost double from the subset (1) with low R$_{Edd}$,  to the subset (4) with high R$_{Edd}$.

The narrow width of Fe~II and H$\beta$ lines, observed in extreme Fe~II emitters, may be due to various physical causes, or due to a combination of some of them. It is most often explained by low black hole mass or by low inclination \citep[see e.g.][]{Marziani2001,Shen2014, Sun2015}. In general, the widths of Fe~II and H$\beta$ can be influenced by certain other factors as well, as  emission region stratification, where dominant emission originates from regions further from the black hole, or by some atomic processes which could produce line narrowing. 
Our results imply that the increase of the R$_{FeII}$ for objects with smaller line widths,  is actually caused by increase in the flux of inconsistent Fe~II lines and  Fe~II$_{ILR}$  components, which are both present in 4434-4684 \AA \ range (see Figure \ref{fig00}). Since ILR components are narrower than VBLR, their larger contribution in total Fe~II flux, leads to narrowing of the profile of Fe~II lines, i.e. to decreasing of the FWHM Fe~II.  Similarly, the relationship between the R$_{FeII}$ parameter and R$_{Edd}$ is actually caused by the relationship of the EW Fe~II$_{ILR}$, and specially the  EW Fe~II$_{incons}$ with R$_{Edd}$ (Figure \ref{fig07}). The question is why the additional Fe~II component (ILR) appears in objects with high Eddington ratio, and in particular, why inconsistent Fe~II lines, which flux should be very weak or negligible according to their transition probabilities, become very strong in these objects.

It is possible that for high Eddington ratio, the additional processes could be activated in the other emission region than one where Fe~II$_{VBLR}$  component arise, which may results with efficient emission of Fe~II$_{ILR}$ and Fe~II$_{incons}$ lines.  Indeed, theoretical considerations of the changes that occur in AGNs with increasing R$_{Edd}$ have estimated that a fundamental change in accretion mode and BLR structure occurs for a critical value of R$_{Edd}$ in the interval 0.1-0.3 \citep{Wang2014,Ganci2019}. In Figure \ref{fig07}, it could be seen that the  highest growth of R$_{FeII}$ is happening for objects with R$_{Edd}$ > 0.1, which belong to the subsets (3) and (4).
This may indicate that fundamental changes in accretion could lead to modifications in the BLR structure and its stratification, as well as to the activation of new processes of Fe~II emission in different layers of the BLR, which would result in  the growth of additional Fe~II components and thus an increase in the total Fe~II emission. 

\begin{figure}
	\centering
\includegraphics[width=75mm]{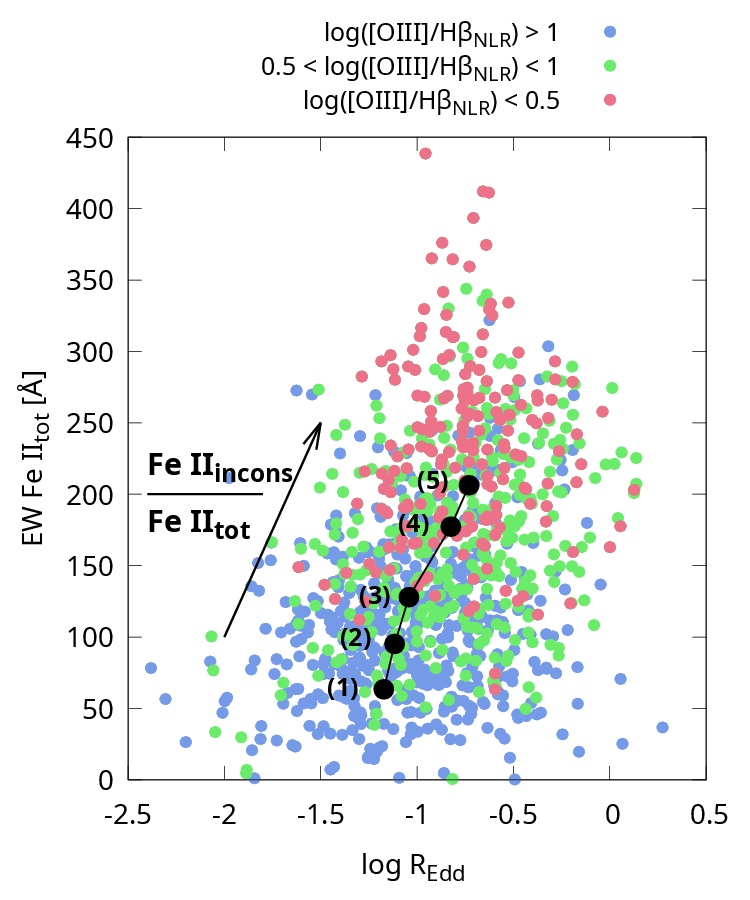}
 \caption{Growth of the contribution of Fe~II$_{incons}$ in total Fe~II, with the increase of the R$_{Edd}$ and decrease of the [O~III]/H$\beta_{NLR}$.  The sample is divided into five subsets according to the Fe~II$_{incons}$/Fe~II$_{tot}$ ratio: (1) Fe~II$_{incons}$/Fe~II$_{tot}$ < 0.1 (87 objects),  (2) 0.1 < Fe~II$_{incons}$/Fe~II$_{tot}$ < 0.2 (134 objects), (3) 0.2 < Fe~II$_{incons}$/Fe~II$_{tot}$ < 0.3 (187 objects), (4) 0.3 < Fe~II$_{incons}$/Fe~II$_{tot}$ < 0.4 (320 objects), and (5) Fe~II$_{incons}$/Fe~II$_{tot}$ > 0.4 (293 objects). For each subset, we found the mean of EW Fe~II$_{tot}$ and  R$_{Edd}$, which are plotted with black dots and indicated by the corresponding numbers (1-5). Additionally, the points are coloured according to log([O~III]/H$\beta_{NLR}$) strength.}
  \label{fig09}
\end{figure}

\subsection{Complexity of the physical processes behind strong Fe~II emission}\label{Sec4.2}

Conventional photoionisation models estimate Fe~II$\lambda$ 4570 \AA /H$\beta$ to be typically 0.2 \citep{Kwan1981},  which is satisfied in the case of objects with weak Fe~II lines. 
In order to theoretically explain the spectra with a significant enhancement in Fe~II emission, the other excitation processes were taken into account, such as collisional excitation, line trapping \citep[see][]{Joly1987, Baldwin2004}, pumping the Fe~II emission by Ly$\alpha$ fluorescence \citep{Marinello2020}, or non-radiative heating due to shocks and outflows \citep{Collin2000}. 
It appears that the processes and physical conditions under which extreme Fe~II emission occurs are still not well understood.
Inclusion of a large number of different atomic processes and data in spectral synthesis code CLOUDY was not sufficient for the theoretical reproduction of the optical Fe~II flux and relative intensities among the Fe~II lines in the case of extremely strong Fe~II emitters \citep{Zhang2024, Pandey2025}. It is suggested that super-solar metallicity in the BLR is required to explain the extreme iron emission \citep{Panda2019b}.  The problem of metallicity in AGN emission regions was dealt by a large number of authors \citep{Hamann2002, Baldwin2003, Negrete2012}. \cite{Gaskell2022} proposed that strong soft X-ray excess increases the iron abundance by orders of magnitude by destroying dust grains and releasing iron into the gas phase. Several studies estimated BLR metallicity of very strong Fe~II emitters by measuring the flux ratios of some broad emission lines \citep{Sniegowska2021,Garnica2022,Floris2024}. As a result, they obtained a super-solar metallicity in the BLR, which reaches several tens of solar metallicity.

In order to explain the high metallicity of BLR required for modelling the extreme Fe~II emission, it has been  proposed that the high metallicity of such objects could be caused by nuclear or circumnuclear star formation activity, which might be associated with supermassive black hole phenomena in some phase of AGN evolution \citep[see][]{Lipari2006, Marziani2025}. In this scenario,  the BLR is consistently enriched in metals by starburst regions that could be located within torus, or even within the high-density gas of the accretion disc \citep{Wang2011,Wang2023, Fan2023, Huang2023}. However, these assumptions are not observationally confirmed since BLR cannot be resolved.

The question is whether the assumption of such a high metallicity is justified or whether the same observational properties could be explained in another way. \cite{Temple2021} found that the several broad emission-line ratios, commonly used in the literature as indicators of BLR metallicity, can be reproduced by a model in which the line emission originates from two kinematically distinct regions with different physical conditions, without requiring metallicities above solar. As pointed out by \cite{Wills1985}, the large observed Fe~II fluxes require either that iron is several times overabundant (compared to solar ratios) or that some unknown process dominates the Fe~II excitation. Here we consider whether the extremely strong Fe~II emission can be explained by the activation of additional atomic processes in certain part of Fe~II emission region.

 Following the Einstein coefficients for spontaneous emission, the optical Fe~II emission is expected to be several orders of magnitude smaller than UV Fe~II, but they are of the same order of magnitude in most of the AGN spectra \citep{Joly1981}. The growth of the optical Fe~II lines relative the UV Fe~II is specially emphasised in strong Fe~II emitters, located at the extreme end of quasar main sequence with R$_{FeII}$>1. It is proposed that mechanism of  self-absorption of the UV Fe~II lines is responsible for this effect, causing the resonantly pumping the optical Fe~II transitions, and therefore their significant increase in AGN spectra \citep{Netzer1980, Joly1981,Collin2000}. However, it seems that similar process is happening among the optical Fe~II lines themselves. The Fe~II$_{incons}$ lines should be up to two orders of magnitude smaller than  Fe~II$_{cons}$, and therefore they are expected to be very weak in spectra, even when Fe~II$_{cons}$ are strong. Since they are of the same order of magnitude as Fe~II$_{cons}$ in strong Fe~II emitters, \cite{Kovacevic2025} proposed that the same mechanism of the self-absorption (radiation trapping) is responsible to transmission of energy from Fe~II$_{cons}$ to Fe~II$_{incons}$ lines.
This effect is possible due to the specific configuration of the energy levels of the Fe~II ions. Namely, the UV Fe~II emission lines at $\sim$2800 \AA \ have the same upper energy level as optical Fe~II emission lines (both, consistent and inconsistent).  At the same time, the lower levels of transitions of UV Fe~II and optical consistent and inconsistent Fe~II lines are all metastabiles (see Grotrian diagram in Figure 12, given in \cite{Kovacevic2025}), which is one of the main reasons for the transfer of energy of radiation first from the UV Fe~II to the optical Fe~II$_{cons}$ lines, and then within optical lines from Fe~II$_{cons}$ to Fe~II$_{incons}$  lines.  Moreover, it is found that both processes are related with growth of the R$_{Edd}$, where it is supposed that a high R$_{Edd}$ causes an increase in the optical depth in the Fe~II lines, which makes the self-absorption process more efficient \citep{Kovacevic2025}.

However, the transmission of energy caused by self-absorption, does not generally explain why, in extreme cases, up to 50\% of the total radiative output from the BLR is released through Fe~II emission \citep{Joly1981}. The observed enhancement of iron emission could be attributed to the formation of localised emission regions within the complex AGN structure, where physical conditions could be suitable for activation of some additional excitation processes. Generally, strong line emission requires efficient population of the upper levels and at the same time efficient depopulation of the lower levels. The efficient population of the upper levels can be achieved by radiative or collisional excitation. For efficient radiative excitation, the shape of the continuum plays an important role, and for efficient collisional excitation, electrons of appropriate kinetic energy are required. 

The strong Eddington ratio is related with appearance of an outflow, which is found to be present in spectra with  strong and narrow optical Fe~II lines \citep{Shen2014}. The strong outflow may cause an effective collisional excitation, which leads to the more efficient population of the upper transition levels of UV and optical Fe~II and at the same time depopulation of their lower metastable levels. 

Our findings suggest that for objects with the strongest Fe~II, the log([O~III]/H$\beta _{NLR}$) is less than 0.5 (see Figure \ref{fig08}), which according to BPT diagram, implies classification into starburst galaxies, LINERs or composite objects (AGN+starburst) \citep[see][]{Kauffmann2003}.  Although this result indicates the possible presence of starburst regions in the NLR, their potential presence closer to the BLR could lead to more efficient collisional excitation and increased gas density, due to shock waves generated by supernova explosions \citep{Lipari2006}, which would lead to enhanced Fe~II emission.

On the other hand, a spectral energy distribution exhibiting enhanced flux at the energies required for ionisation and excitation of iron would make the photoexcitation of Fe~II ions more efficient.  The ionisation and excitation energies for optical Fe~II are relatively low (7.9 eV and $\sim$5 eV, respectively), and their sum  in units of wavelength corresponds to $\sim$ 960 \AA. Therefore, the softer spectral energy distribution, with strong radiation in far-UV would contribute to a more efficient excitation of the upper levels of the optical Fe~II lines. Indeed, several theoretical models of spectral energy distribution predict a softening of the ionising continuum and more luminous far-UV continuum for the sources with high Eddington ratio (R$_{Edd}$>0.1), compared to those with lower Eddington ratio \citep{Kubota2018,  Ferland2020, Garnica2025}. Moreover, the observed variation of the [O~III]/H$\beta _{NLR}$ ratio along the quasar main sequence (Figure \ref{fig08}) can be interpreted as a change in the shape of the ionising continuum within the NLR \citep{Baldwin1981}. The formation of [O~III] ion requires an energy of 48.7 eV (the sum of the first and second ionisation energies), while a total energy of 12.7 eV is required to excite the upper level of the H$\beta$ transition. The low ratio of the [O~III]/H$\beta _{NLR}$ in strong Fe~II emitters indicates that in NLR of these objects, the excited H$\beta$ atoms are more likely to be formed than double ionised [O~III] ions, which require higher energies. If we assume that the shape of the ionising continuum in BLR is similar as the ionising continuum in NLR, it could imply presence of a softer ionising continuum in BLR of strong Fe~II emitters, comparing to weak Fe~II emitters, which would be very favourable for the excitation of the upper levels of the considered optical Fe~II lines.  This assumption is observationally supported since in the spectra of strong Fe~II emitters, the lines of the helium ion (such as He II 4686 \AA) tend to be smaller relative to the Fe~II lines compared to those in weaker Fe~II sources. Note that energy of helium ionisation is 24.6 eV, and energy of excitation of He II 4686 \AA \ from the ground state is 51 eV. 

 However, the decrease in the [O~III]/H$\beta _{NLR}$ ratio may also be caused by other physical conditions. An increase in the electron density of the NLR towards the critical density of [O~III] ($\sim$10$^5$ cm$^{-3}$) would lead to more efficient collisional de-excitation, suppressing the [O~III] emission and consequently  decreasing the [O~III]/H$\beta _{NLR}$ ratio \citep{Osterbrock2006, Wolf2026}. Another possible explanation is the weakening of [O~III] due to the so-called disappearing NLR in high-luminosity sources \citep{Netzer2004}. In addition, there are indications that metallicity may also affect the [O~III]/H$\beta _{NLR}$ ratio \citep{Dopita2006, Feltre2016}.

 In Figure \ref{fig09} we summarised the mentioned properties of optical Fe~II emission. 
  It could be seen that total Fe~II emission grows as contribution of the Fe~II$_{incons}$ grows in total Fe~II flux, which is related with increase of R$_{Edd}$ and decrease of [O~III]/H$\beta _{NLR}$ ratio.  Similarly as noticed for narrow line width, the high Eddington ratio is a necessary, but not a sufficient condition for very strong Fe~II emission. On the other hand, the [O~III]/H$\beta _{NLR}$ ratio appears to be  better indicator of extreme Fe~II emitters, with low values of this ratio reflecting the particular physical conditions that characterise this subpopulation.

 \section{Conclusions}\label{Sec5}
 
 In this study, we have used a large sample of Type 1 AGN spectra to investigate the complex nature of the Fe~II emission region by analysing correlations between Fe~II features and other spectral parameters. We focused in particular on the quasar main sequence anti-correlation to better understand the underlying physics, especially in the case of strong Fe~II emitters. To achieve this, we applied a flexible and complex Fe~II template that enabled us to decompose the Fe~II emission in the optical range and to follow the behaviour of different groups of Fe~II lines and their components along the quasar main sequence.  The optical Fe~II lines were divided into two large line groups, referred to as consistent and inconsistent, based on their transition probabilities. Furthermore, for the consistent Fe~II lines, we applied a two-component model and separately examined the relative contributions from the ILR and the VBLR within the framework of the quasar main sequence. Summarising the main results, we outline the following conclusions:
 
\begin{enumerate} 
\item The Fe~II  lines with widths below $\sim$ 5000 km s$^{-1}$ can be satisfactorily reproduced using a two-component model. In this model, the line wings are fitted with a broader Gaussian component associated with emission from the VBLR, while the line cores are fitted by a narrower Gaussian corresponding to emission from the ILR. The applied two-component Fe~II model provides an improved accuracy in fitting Fe~II lines in these objects. However, for sources exhibiting broader Fe~II profiles (FWHM larger than $\sim$5000 km s$^{-1}$), the model does not significantly enhance the fit quality, implying that in these objects the Fe~II emission likely originates from a single kinematic region.

\item The anti-correlation between the EW Fe~II  and FWHM Fe~II  appears to represent a more fundamental relation underlying the quasar main sequence. Replacing the H$\beta$ FWHM with the Fe~II FWHM, obtained applying the complex Fe~II template, significantly strengthens the quasar main sequence anti-correlation. 

\item  The increase in EW Fe~II (and consequently in the parameter R$_{FeII}$) along the quasar main sequence is primarily caused by the enhancement of EW Fe~II$_{incons}$ lines and, with a smaller contribution, by the enhancement of EW Fe~II$_{ILR}$  components.
The relative contribution of these components to the total Fe~II flux increases with increasing Eddington ratio and decreasing Fe~II line width. 
In contrast, the EW of Fe~II$_{VBLR}$  components remain, on average, similar in both strong and weak Fe~II emitters, exhibiting no significant variation along the quasar main sequence.

\item The strong Fe~II emitters (EW Fe~II > 150 \AA) typically exhibit R$_{Edd}$ > 0.1, FWHM Fe~II < 5000 km s$^{-1}$, and log[O~III]/H$\beta _{NLR}$ < 1. In these objects, the Fe~II$_{incons}$  and Fe~II$_{ILR}$  components together contribute more than 50\% of the total Fe~II flux. However, small line widths and a high Eddington ratio appear to be necessary but not sufficient conditions for extreme Fe~II emission, while the [O~III]/H$\beta _{NLR}$ ratio represents a more reliable diagnostic of such objects. Additionally, these objects exhibit a different slope in the FWHM Fe~II  - L$_{5100}$ relation relative to the rest of the sample, implying a distinct physical origin and pointing towards an underlying AGN dichotomy.
 
\item   The results indicate a possible stratification of the Fe~II emission region occurring in sources with very strong Fe~II emission. 
 A high Eddington ratio may induce structural changes in the disc and BLR, resulting in a softening of the ionising continuum or the appearance of strong outflows that enhance collisional excitation. This can lead to the development of localised emission regions with specific physical conditions favourable to activating additional atomic processes. These processes could be responsible for the increase of the Fe~II$_{incons}$ lines and Fe~II$_{ILR}$ components and consequently for the enhancement of the observed Fe~II strength.

\end{enumerate} 

Future research should be devoted to a more careful consideration of the possible atomic processes and physical conditions that could be responsible for the emission of additional Fe~II components. Understanding the origin of the enhanced Fe~II emission is essential for gaining better insight into the complex physics and structure of the BLR.
 
\begin{acknowledgements}
We thank the anonymous referee for valuable comments
and suggestions that helped us to significantly improve the paper. This research was supported by the Ministry of Science, Technological Development and Innovation of the Republic of Serbia through contracts no. 451-03-33/2026-03/200002 and 451-03-33/2026-03/200162.    

Funding for the Sloan Digital Sky Survey IV has been provided by the 
Alfred P. Sloan Foundation, the U.S. 
Department of Energy Office of 
Science, and the Participating 
Institutions. 

SDSS-IV acknowledges support and 
resources from the Center for High 
Performance Computing  at the 
University of Utah. The SDSS 
website is www.sdss4.org.

SDSS-IV is managed by the 
Astrophysical Research Consortium 
for the Participating Institutions 
of the SDSS Collaboration including 
the Brazilian Participation Group, 
the Carnegie Institution for Science, 
Carnegie Mellon University, Center for 
Astrophysics | Harvard \\\& 
Smithsonian, the Chilean Participation 
Group, the French Participation Group, 
Instituto de Astrof\'isica de 
Canarias, The Johns Hopkins 
University, Kavli Institute for the 
Physics and Mathematics of the 
Universe (IPMU) / University of 
Tokyo, the Korean Participation Group, 
Lawrence Berkeley National Laboratory, 
Leibniz Institut f\"ur Astrophysik 
Potsdam (AIP),  Max-Planck-Institut 
f\"ur Astronomie (MPIA Heidelberg), 
Max-Planck-Institut f\"ur 
Astrophysik (MPA Garching), 
Max-Planck-Institut f\"ur 
Extraterrestrische Physik (MPE), 
National Astronomical Observatories of 
China, New Mexico State University, 
New York University, University of 
Notre Dame, Observat\'ario 
Nacional / MCTI, The Ohio State 
University, Pennsylvania State 
University, Shanghai 
Astronomical Observatory, United 
Kingdom Participation Group, 
Universidad Nacional Aut\'onoma 
de M\'exico, University of Arizona, 
University of Colorado Boulder, 
University of Oxford, University of 
Portsmouth, University of Utah, 
University of Virginia, University 
of Washington, University of 
Wisconsin, Vanderbilt University, 
and Yale University.
\end{acknowledgements}


\begin{appendix}

\section{Two-component model of Fe~II lines}\label{A}

The two-component model of the broad emission lines, which assumes that the BLR consists of two subregions (ILR and VBLR), has been considered in many studies
\citep[see][and references therein]{Hu2008, Li2015}. The ILR is assumed to be further away from the black hole comparing the VBLR, with line widths  FWHM $\sim$ 2000 km s$^{-1}$ \citep{Brotherton1994,Li2015}, while the comparison of the widths of ILR and VBLR components in H$\beta$ gives approximately two to three times wider VBLR components \citep{Hu2008,Kovacevic2010,Kovacevic2015}. 

It is found that the Fe~II lines are kinematically connected to the H$\beta$ ILR components \citep{Hu2008,Kovacevic2010}, which implies their ILR origin, while \cite{Kovacevic2010} indicate that it is also possible that part of the Fe~II emission originates from VBLR. On the other hand, several empirical studies of optical Fe~II lines \citep{Veron-Cetty2004,Dong2008,Bruhweiler2008,Park2022}, noticed that Fe~II lines have a broader and narrower components in some spectra \citep[see Appendix B in][]{Kovacevic2025}. \cite{Popovic2023} showed that the set of model spectra constructed as different rates of the ILR/VBLR contribution in H$\beta$ and Fe~II lines could reproduce the quasar main sequence. This motivated us to apply a two-component Fe~II model for this research where consistent Fe~II lines are fitted with ILR and VBLR components that are not tied to the same components of the H$\beta$.

However, applying the two-component Fe~II model to an already complex Fe~II template with several free parameters raises questions about the reliability and uniqueness of the fit decomposition, especially in spectra with very broad Fe~II lines. Similar to H$\beta$, the ILR component of the Fe~II line is expected to be narrower than the VBLR component and to have a FWHM of approximately $\sim$ 2000 km s$^{-1}$, implying that the two-component model is not justified for spectra with very broad Fe~II lines. Namely, in these spectra there are no clearly visible peaks of the Fe~II lines, because the lines are very broad and overlap, forming smooth bumps, which do not show any indication of the existence of two distinct emission regions.

\begin{figure}
	\centering
	\includegraphics[width=60mm, angle=270]{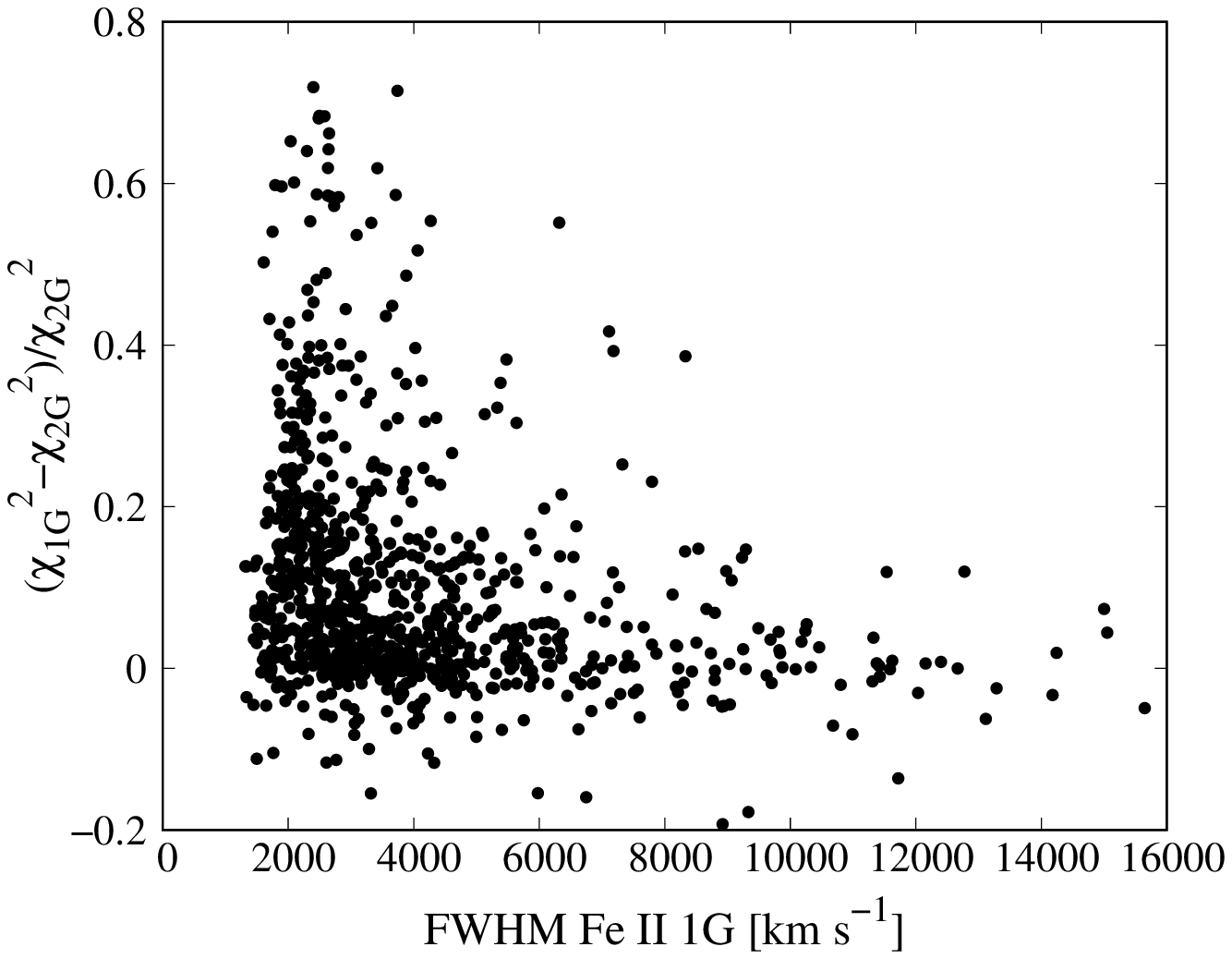}
	      \caption{Comparison of the quality of fit of the Fe~II$_{cons}$ lines with a single-Gaussian and two-Gaussian model (ILR+VBLR). On the Y-axis is the normalised difference between $\chi^2$ of fit of Fe~II$_{cons}$ with a single-Gaussian model ($\chi^2_{1G}$) and two-Gaussian model ($\chi^2_{2G}$).  On the X-axis is the  Fe~II$_{cons}$ width, estimated using single-Gaussian model (FWHM Fe~II$_{1G}$).	   }
    \label{fig001}
\end{figure}

 We tested the validity of using the two-component Fe~II model for fitting the sources with different Fe~II line widths by comparing the $\chi^2$ values obtained with the two-component Fe~II fit to those derived from the single-component Fe~II fit. Both models were applied to the full sample, and the resulting $\chi^2$ differences were plotted as a function of the Fe~II line width estimated with the single-component Fe~II model (FWHM Fe~II$_{1G}$) (see Fig. \ref{fig001}).  It can be seen that, for large number of objects with narrower Fe~II lines (FWHM Fe~II$ _{1G}$ < 5000 km s$^{-1}$), the Fe~II fit is significantly improved with the two-component Fe~II model.  On the other hand, 
  for most of objects with very broad Fe~II lines (FWHM Fe~II$_{1G}$ > 5000 km s$^{-1}$), adopting a two-component Fe~II model does not lead to a significant improvement in the fit quality compared to a single-component model. Additionally, in approximately 37\% of objects with FWHM Fe~II$_{1G}$ > 5000 km s$^{-1}$, the two-component Fe~II model is reduced to a single-component one, as the $\chi^2$ minimisation routine discarded one of the components to achieve the best fit. Since for these objects the decomposition is identical to the single-component one, they are not included in Fig. \ref{fig001}. 
 
Therefore, we found that application of the two-component Fe~II model is justified for spectra with Fe~II widths smaller than FWHM Fe~II$ _{1G}$ $\sim$ 5000 km s$^{-1}$. On the other hand, for spectra with very broad Fe~II lines FWHM Fe~II$ _{1G}$ > 5000 km s$^{-1}$, fitting with the Fe~II two-Gaussian model is unreliable because it does not have a unique decomposition and does not improve the quality of the fit compared to the single-Gaussian model.

\section{Distribution of the sources with different properties in the log$\lambda L_{5100}$ versus FWHM Fe~II diagram}\label{B}
 
\begin{figure}
	\centering
	\includegraphics[width=60mm, angle = 270]{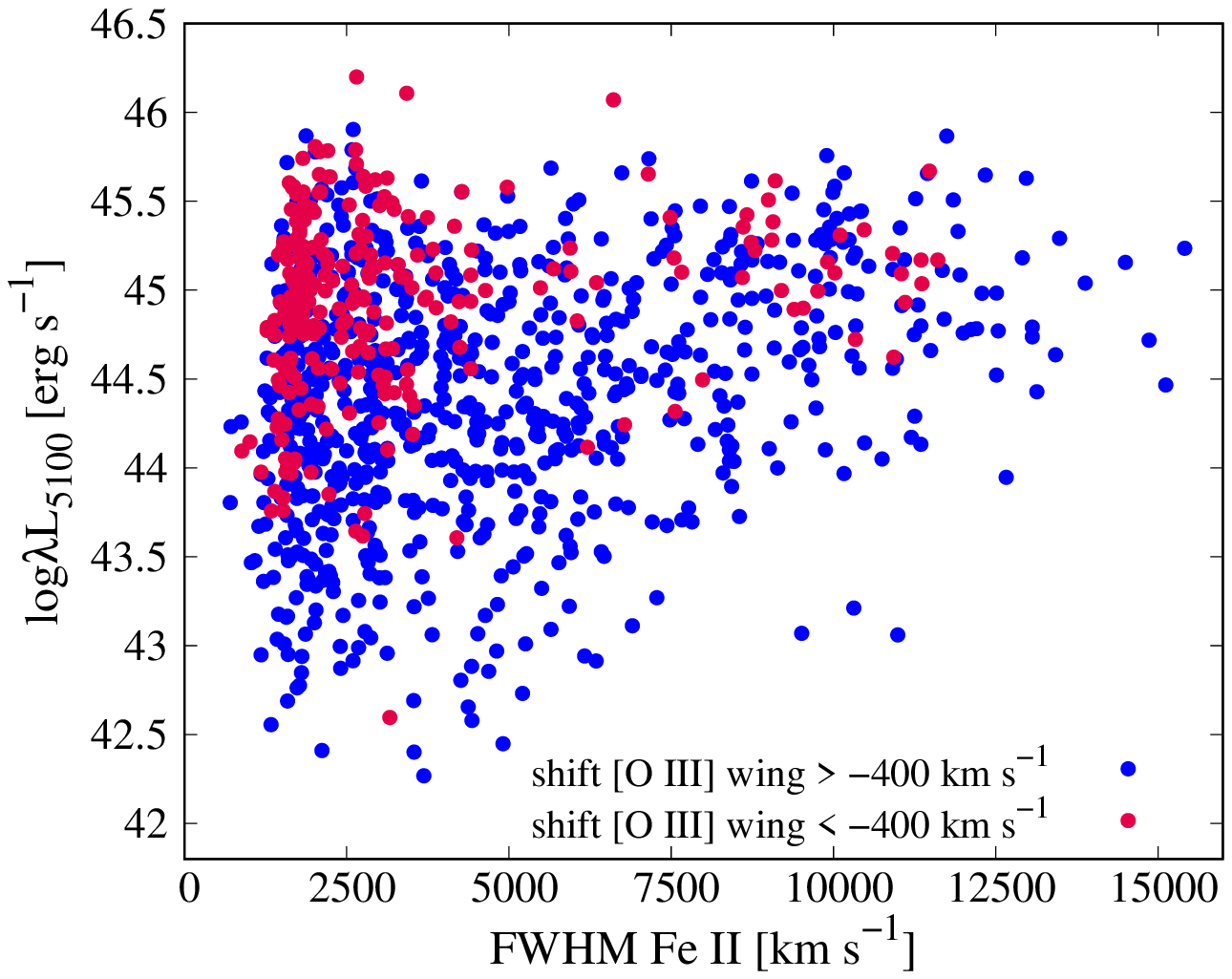}
	 \caption{Outflow signatures in the log$\lambda L_{5100}$ versus FWHM Fe~II  plane. Red and blue points indicate sources with [O~III] wing  component shifts smaller and larger than -400 km s$^{-1}$, respectively.}
    \label{B1}
\end{figure}

  Since the gas outflows can play an important role in triggering different atomic processes,  we examined the distribution of the outflow signatures within the  log$\lambda L_{5100}$ versus FWHM Fe~II diagram. Previous studies have shown that the wing component of the [O~III] 5007 \AA \  emission line is predominantly governed by non-gravitational outflow kinematics, while the [O~III] core component traces both gravitational and outflow kinematics \citep{Woo2016, Sexton2021,Kovacevic2022}. 
In our fitting procedure, the [O~III] lines are modelled with wing and core components. To trace the outflow kinematics, we used the shift of the [O~III] wing components corrected for systemic shift of the narrow Balmer lines.  We found that [O~III] wing components with the largest blueshifts (< - 400 km s$^{-1}$) are located in the left branch of the diagram (see Figure \ref{B1}). 

\begin{figure}
	\centering
	\includegraphics[width=60mm, angle = 270]{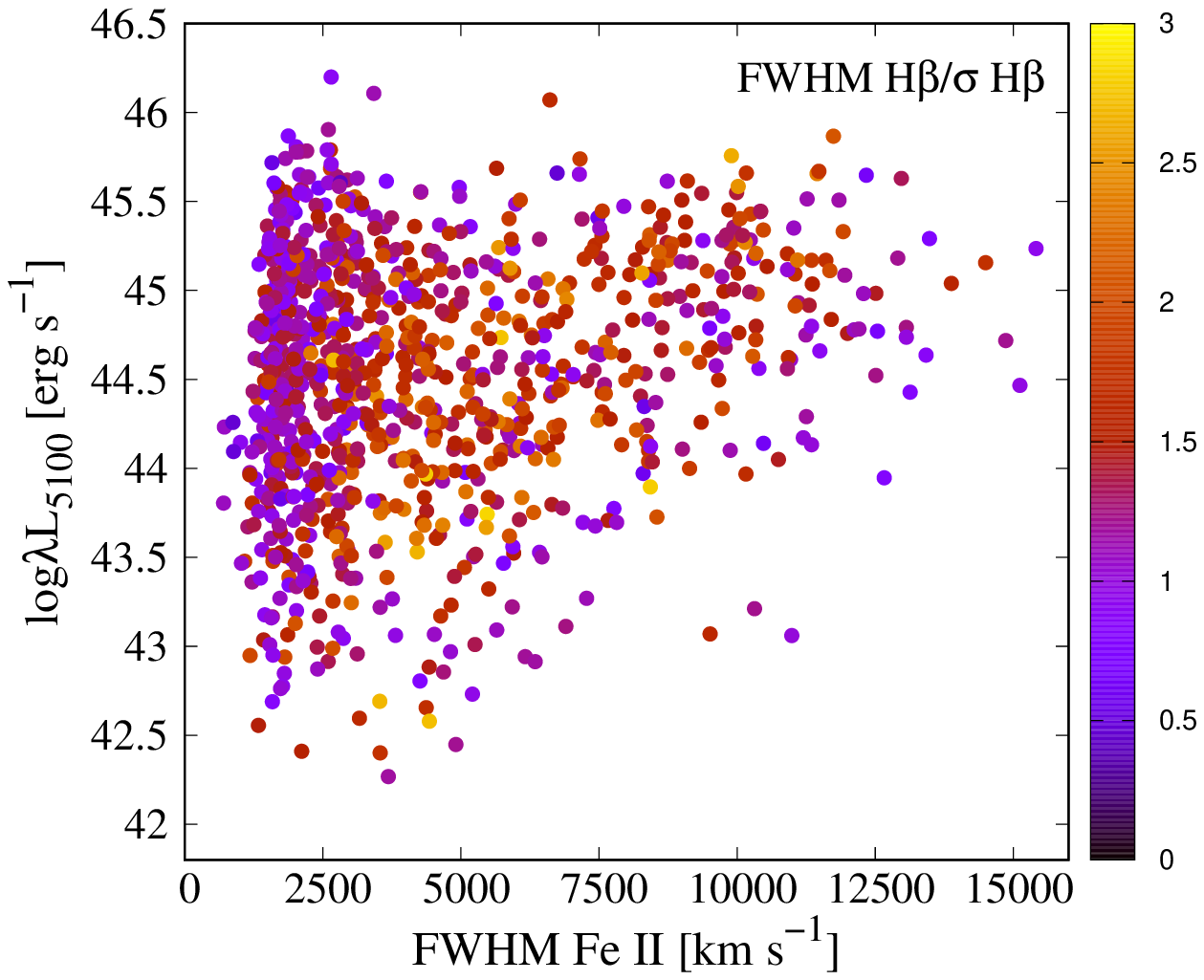}
 \caption{Variation of the FWHM H$\beta$/$\sigma$H$\beta$ ratio across the log$\lambda L_{5100}$ versus FWHM Fe~II  parameter space. The colour scale indicates the value of the FWHM H$\beta$/$\sigma$H$\beta$ ratio. }
    \label{B2}
\end{figure}

We analysed the changes in the shape of the broad H$\beta$ line across the log$\lambda L_{5100}$ versus FWHM Fe~II plane by following the variations in the ratio of FWHM H$\beta$ to the line dispersion of the total  H$\beta$ profile (FWHM H$\beta$/$\sigma$ H$\beta$). Our results show that objects with lower values of this ratio are predominantly located in the left branch of the diagram (see Figure \ref{B2}). This may indicate differences in inclination or in BLR structure between objects from two branches.

\end{appendix}

\end{document}